\documentclass[journal=jctcce,manuscript=article]{achemso}   

\usepackage[version=3]{mhchem} 
\usepackage{multirow}
\usepackage{soul}
\usepackage{array}
\usepackage[usenames,dvipsnames]{color} 
\usepackage[utf8]{inputenc}
\usepackage[T1]{fontenc}
\PassOptionsToPackage{hyphens}{url}
\usepackage{hyperref,url}
\usepackage{mathtools}
\usepackage{bm}
\usepackage{braket}
\usepackage{calrsfs}
\usepackage{multicol}
\usepackage{multirow}
\usepackage{makecell} 
\definecolor{mygreen}{rgb}{0.2,0.7,0.2}
\usepackage{dcolumn}
\usepackage{amsmath}
\usepackage{latexsym}
\usepackage{subcaption}
\usepackage{graphicx}
\usepackage{amssymb}
\usepackage{lineno} 
\usepackage{float}
\usepackage{physics}
\DeclareMathAlphabet{\pazocal}{OMS}{zplm}{m}{n}
\usepackage{bbm}

\newcommand{\bluerow}[3]{\textcolor{blue}{#1}&\textcolor{blue}{#2}&\textcolor{blue}{#3}}
\newcommand{\redrow}[3]{\textcolor{red}{#1}&\textcolor{red}{#2}&\textcolor{red}{#3}}

\author{M. A. Garc\'ia-Bl\'azquez}
\email{manuelantonio.garcia@estudiante.uam.es}
\affiliation{~Departamento de F\'\i sica de la Materia Condensada, Universidad Aut\'onoma de Madrid, E-28049 Madrid, Spain}

\author{J. J. Esteve-Paredes}
\affiliation{~Departamento de F\'\i sica de la Materia Condensada, Universidad Aut\'onoma de Madrid, E-28049 Madrid, Spain}

\author{A. J. Uría}
\affiliation{~Departamento de F\'\i sica de la Materia Condensada, Universidad Aut\'onoma de Madrid, E-28049 Madrid, Spain}

\author{J. J. Palacios}
\affiliation{~Departamento de F\'\i sica de la Materia Condensada, Universidad Aut\'onoma de Madrid, E-28049 Madrid, Spain}
\alsoaffiliation{Condensed Matter Physics Center (IFIMAC), Universidad Autónoma de Madrid, E-28049 Madrid, Spain}

\title{Shift current with Gaussian basis sets \& general prescription for maximally-symmetric summations in the irreducible Brillouin zone}

\begin{document}
\graphicspath{{Images/}}

\begin{tocentry}
\includegraphics[width=8.25cm,height=4.45cm]{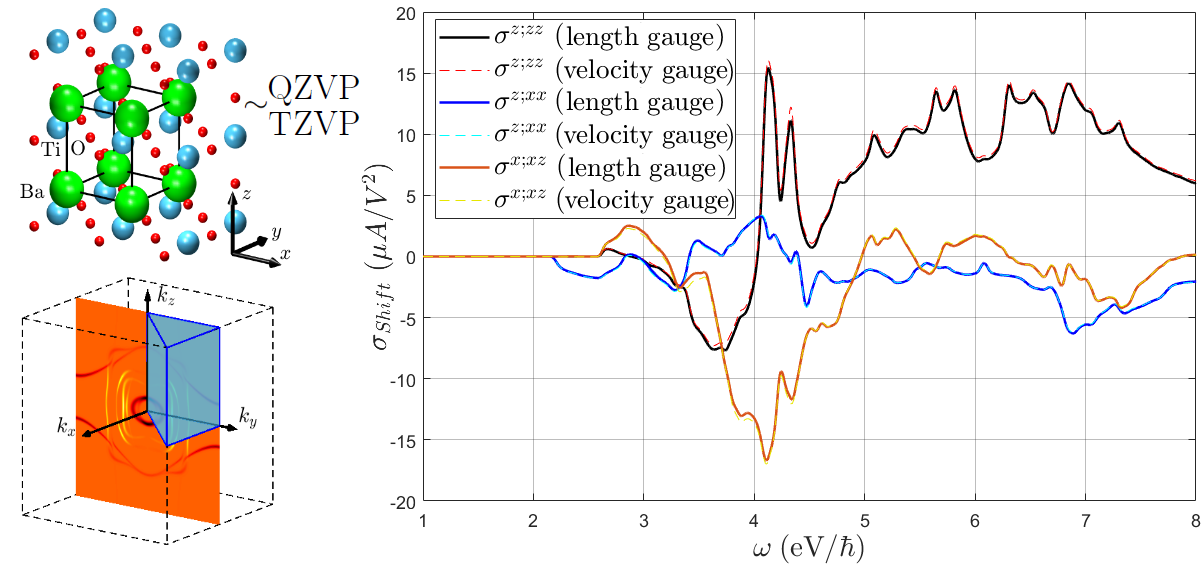}
\end{tocentry}


\begin{abstract} 
The bulk photovoltaic effect is an experimentally verified phenomenon by which a direct charge current is induced within a non-centrosymmetric material by light illumination. Calculations of its intrinsic contribution, the shift current, are nowadays amenable from first-principles employing plane-waves bases. In this work we present a general method for evaluating the shift conductivity in the framework of localized Gaussian basis sets that can be employed in both the length and velocity gauges, carrying the idiosyncrasies of the quantum-chemistry approach. The (possibly magnetic) symmetry of the system is exploited in order to fold the reciprocal space summations to the representation domain, allowing to reduce computation time and unveiling the complete symmetry properties of the conductivity tensor under general light polarization.    
\end{abstract}

\flushbottom
\maketitle

\thispagestyle{empty}

\section{Introduction}
The generation of a non-oscillating response in a material medium under an incident electric oscillating field is a general feature that occurs at all even orders in the perturbative expansion, that is, it is a non-linear optical phenomenon. For responses that transform as vectors, such as an electric current, an elemental symmetry analysis shows that all the even order response tensors must vanish in the presence of inversion symmetry, hence such frequency-independent quantity can only arise in non-centrosymmetric materials. In this regard, the emergence of a direct charge current in an homogeneous material induced by light is known as the bulk photovoltaic effect (BPVE) \cite{sturman1992photovoltaic}. It was established by the mid seventies with earlier experimental reports in ferroelectric materials \cite{chynoweth1956surface,chen1969optically,glass1974appl,koch1975bulk}, and continued gathering attention during the next decades \cite{fridkin1977effect,kraut1979anomalous,von1981theory,hornung1983band,fridkin1993bulk,batirov1997bulk,buse1997light,kral2000photogalvanic,Sipe2000}. However, the potential applications in solar cells \cite{butler2015ferroelectric,spanier2016power}, and the advances in both experimental facilities and first-principles capabilities have driven a considerable surge of studies in recent years \cite{dai2023recent,cook2017design,rangel2017large,osterhoudt2019colossal,Nagaosa2020,wang2020polarization,BSPVE,martinez2022direct,chaudhary2022shift,zhang2022tailoring}. 

The BPVE is part of the total second-order optical response, which in addition includes second-harmonics and, for polychromatic electric fields, contributions of mixed frequencies. In turn, the BPVE can be separated into 3 essentially different contributions \cite{dai2023recent}: the shift current, a static and coherent (stemming from the off-diagonal part of the density matrix) response that under time-reversal ($\hat{\pazocal{T}}$) symmetry appears only with linearly polarized light; and two transient contributions that eventually reach a steady state, namely the injection current, which under $\hat{\pazocal{T}}$ symmetry appears only with circularly polarized light, and the ballistic current, which under $\hat{\pazocal{T}}$ symmetry emerges purely from coherent scattering processes such as electron-phonon or electron-hole interactions that introduce an imbalance between the carrier generation rates across the Brillouin zone (BZ). Disregarding excitonic effects, the BPVE in non-metallic systems occurs at frequencies above the band gap. These quantities, or equivalently the corresponding third-rank tensors $\sigma^{a;bc}(\omega)$ as a function of a single frequency, admit expressions in terms of the quasi-particle properties that are are amenable to numerical evaluation via quantum mechanical methods. Specifically, these microscopic expressions can be obtained by diagrammatic approaches for the ballistic current \cite{dai2021phonon,xu2022nonlinear,dai2021first}, and solving the density matrix perturbatively \cite{sturman1992photovoltaic}, employing Wilson loops \cite{wang2022generalized} or again by diagrammatic techniques \cite{Moore2019} for the injection and shift currents. However, only the latter one is truly intrinsic to the single-particle system, in the sense that it can be computed exclusively from the band structure and electronic eigenfunctions without further modelling. 

The calculation of the shift current presents some difficulties or subtleties starting from the choice of gauge for the interaction of electrons with the field \cite{ventura2017gauge,Pedersen2017,passos2018nonlinear,Moore2019}. The most generally applicable method, the length gauge, requires evaluating numerical derivatives with respect to the crystalline momentum $\bm{k}$ of quantities that are not gauge invariant. In contrast, the velocity gauge constitutes a more straightforward alternative, although it carries an (a priori) infinite sum over the electronic states external to the direct optical transition. Both gauges require the evaluation of the matrix representation of the velocity operator in the set of crystalline eigenfunctions, and both are expected to yield equal results in the limit of a complete basis for describing the latter. There currently exist methods for evaluating the shift current in the single-particle approximation from density-functional theory (DFT) \cite{Rappe2012}, tight-binding (including Wannierizations \cite{wang2017first,Ivo2018}) and $\bm{k}\cdot\bm{p}$ band structures \cite{cook2017design}. Yet, as it is often the case in physics-leaning studies, the DFT calculations are almost invariably assumed to employ a plane-wave basis, at least when the velocity operator is not approximated by the momentum. 

In this work, we present a formalism for computing the shift conductivity tensor in non-metallic crystals, in both length and velocity gauges, from first-principles employing Gaussian basis sets. It is based on an exact calculation of the velocity and Berry connection matrix elements through the analytical evaluation of the real-space integrals involved. The use of a localized basis presents some advantages and disadvantages with respect to the plane-waves alternative inherited from the DFT methods: the whole chain of calculations should generally be faster, the evaluation of position (and by extension, velocity) matrix elements is straightforward, hybrid functionals can be used at little cost (which may allow to obtain an accurate gap avoiding scissor corrections or GW calculations), all-electron calculations can be performed, and no artificial replication of layers is required in 2D materials. On the other hand, a customized basis optimization has to be performed for each system while limited by the superposition error and diffusive exponents, and errors from the lack of completeness of the basis are more likely (making the more delocalized unoccupied states particularly hard to reproduce). It is expected that the reliability of this method is highly correlated with the ability to properly reproduce the relevant occupied and unoccopied states (dictated by the frequency range) with a Gaussian basis. 

A further benefit of the use of localized bases lies in the guarantee that the complete symmetry of the system is preserved, in slight contrast with the maximally-localized Wannier representation. The properties of the crystallographic point group can then be exploited to reduce the summations over the BZ that are required for the shift conductivity to properly weighted sums, which encode the whole (possibly magnetic) symmetry of the system, only over its irreducible part or representation domain. We present a complete list of the explicit formulae for each space group including the magnetic configurations, where the structure type is used to parametrize the irreducible domain and the (magnetic) point group determines the precise folding of the $\bm{k}-$resolved conductivity. 

\section{Shift current: definition, considerations \& numerical evaluation}
A general expression for the total second-order optical response under homogeneous illumination can be obtained by solving the density matrix in perturbation theory for the field. In particular, for a uniform polychromatic electric field $\tilde{\bm{E}}(t)=\sum_{j}\bm{E}(\omega_{j})e^{-i\omega_{j} t}+\text{c.c.}$, the shift current is defined as the intrinsic second-order DC component 
\begin{equation}\label{current}
J^{a}_{\text{shift}}=2\sum_{j\:\vert\:\omega_{j}>0}\sum_{b,c}\Re\left[\sigma^{a;bc}_{\text{shift}}(\omega_{j})E^{b}(\omega_{j})E^{c}(-\omega_{j})\right]
\end{equation}
where $a,b,c$ label the spatial components in the chosen coordinate system.

\subsection{Length gauge}\label{secLG}
In the length gauge, the electric potential is chosen as $\hat{V}=\abs{e}\hat{\bm{r}}\cdot\bm{E}$ and the perturbative expression for the shift conductivity third-rank tensor in a non-metallic material ultimately reads \cite{AversaSipe,Sipe2000,Nagaosa2020}
\begin{equation}\label{SCgd1} 
\sigma^{a;bc}_{\text{shift}}(\omega)=-\frac{ig_{s}\pi\abs{e}^3}{2\hbar^{2}V}\sum_{\bm{k}\in\text{BZ}}\sum_{m,n} 
f_{m,n}(\bm{k})\left[ A^{b}_{m,n}(\bm{k})A^{c}_{n,m;a}(\bm{k}) - A^{c}_{n,m}(\bm{k})A^{b}_{m,n;a}(\bm{k}) \right]\delta(\omega_{m,n}(\bm{k})-\omega)
\end{equation}
which is valid irrespective of whether time-reversal $\hat{\pazocal{T}}$ is a symmetry. In this formula
\begin{itemize}
\item $m,n$ label the eigenstates of the periodic single-particle Hamiltonian $\hat{H}\ket{\psi_{n,\bm{k}}}=E_{n,\bm{k}}\ket{\psi_{n,\bm{k}}}$, which satisfy Bloch's theorem: $\bra{\bm{r}}\ket{\psi_{n,\bm{k}}}=\psi_{n,\bm{k}}(\bm{r})=e^{i\bm{k}\bm{r}}u_{n,\bm{k}}(\bm{r})$ with $u_{n,\bm{k}}$ having the periodicity of the direct lattice. $\bm{k}$ is the crystalline momentum or label of the irreducible representations of the translation group. $\hbar\omega_{m,n}(\bm{k})\equiv E_{m,\bm{k}}-E_{n,\bm{k}}$ and $f_{m,n}(\bm{k})\equiv f_{m}(\bm{k})-f_{n}(\bm{k})$ is the difference of Fermi distributions, hereafter taken at zero temperature. 
\item $A^{b}_{m,n}(\bm{k})=\left.\bra{u_{m,\bm{k}'}}i\hat{\partial}_{k^{\prime b}}\ket{u_{n,\bm{k}'}}\right\vert_{\bm{k}}$ is the $b-$th spatial component of the Berry connection matrix elements, which satisfy $A^{b}_{m,n}(\bm{k})=A^{b}_{n,m}(\bm{k})^{*}$.
\item $A^{b}_{n,m;a}(\bm{k})=\left.\partial_{k^{\prime a}}A^{b}_{n,m}(\bm{k}')\right\vert_{\bm{k}}-i[A^{a}_{n,n}(\bm{k})-A^{a}_{m,m}(\bm{k})]A^{b}_{n,m}(\bm{k})$ is the generalized derivative (GD) of the Berry connection. 
\item $V=N_{\bm{k}}V_{\text{PUC}}$ is the volume (area in 2D, or longitude in 1D) of the crystal, with $N_{\bm{k}}\to\infty$ the number of terms in the Brillouin zone (BZ) summation (or discretized integration) and $V_{\text{PUC}}$ the volume of the primitive unit cell. $\delta(\omega)$ is a nascent (or broadened) Dirac delta function. $g_{s}=1$ (2) in the presence (absence, respectively) of spin-dependent terms in the Hamiltonian (excluding the doubled states in the $m,n$ summations).
\item We have introduced a global minus sign in agreement with Ref. \cite{Nagaosa2020} and equation 58 of Ref. \cite{Sipe2000}. We note that some authors do not include the $1/2$ factor in the conductivity, instead cancelling it with the prefactor in \eqref{current}.
\end{itemize}

It follows that in general $\sigma^{a;bc}_{\text{shift}}(\omega)=\sigma^{a;cb}_{\text{shift}}(\omega)^{*}=\sigma^{a;bc}_{\text{shift}}(-\omega)^{*}$ and $\sigma^{a;bb}_{\text{shift}}$ is real. For linearly polarized light, $E^{b}(\omega_{j})E^{c}(-\omega_{j})$ is real for all components and only $\Re\sigma^{a;bc}_{\text{shift}}$ contributes to $J^{a}_{\text{shift}}$. In contrast, for circular polarization $E^{b}(\omega_{j})E^{c}(-\omega_{j})$ is complex for some $b,c$, hence both the real and imaginary parts of the conductivity may contribute to the current in general. Under $\hat{\pazocal{T}}$ symmetry, in particular excluding any permanent magnetic alignment, $\hat{\pazocal{T}}u_{n,\bm{k}}=e^{i\theta_{n,\bm{k}}}u_{n,-\bm{k}}$ for an arbitrary phase $\theta_{n,\bm{k}}$; hence applying the anti-unitary transformation in the inner products, and noting that $\hat{\pazocal{T}}i\hat{\bm{\nabla}}_{\bm{k}}\hat{\pazocal{T}}^{-1}=-i\hat{\bm{\nabla}}_{\bm{k}}$
\begin{equation*}\begin{aligned}
&A^{b}_{m,n}(\bm{k})=
\left.\bra{u_{m,-\bm{k}'}}e^{-i\theta_{m,\bm{k}}}-i\hat{\partial}_{k^{\prime b}}e^{i\theta_{n,\bm{k}}}\ket{u_{n,-\bm{k}'}}^{*}\right\vert_{\bm{k}}=e^{i\theta_{n,m}^{\bm{k}}}\left[A^{b}_{m,n}(-\bm{k})^{*}+\frac{\partial\theta_{n,\bm{k}}}{\partial k^{b}}\delta_{m,n}\right]\:\Rightarrow \\
&A^{c}_{n,m;a}(\bm{k})=-e^{i\theta_{m,n}^{\bm{k}}}\left[ \left.\frac{\partial A^{c}_{n,m}(\bm{k}')^{*}}{\partial k^{\prime a}}\right\vert_{-\bm{k}} +i\left(A^{a}_{n,n}(-\bm{k})^{*}-A^{a}_{m,m}(-\bm{k})^{*}\right)A^{c}_{n,m}(-\bm{k})^{*} \right]
=-e^{i\theta_{m,n}^{\bm{k}}}A^{c}_{n,m;a}(-\bm{k})^{*}
\end{aligned}\end{equation*}
where $\theta_{n,m}^{\bm{k}}\equiv\theta_{n,\bm{k}}-\theta_{m,\bm{k}}$. Thus $f_{m,n}(\bm{k})A^{b}_{m,n}(\bm{k})A^{c}_{n,m;a}(\bm{k})=-\left[f_{m,n}(-\bm{k})A^{b}_{m,n}(-\bm{k})A^{c}_{n,m;a}(-\bm{k})\right]^{*}$ and $\sigma^{a;bc}_{\text{shift}}$ is real with $\hat{\pazocal{T}}$ symmetry. In this case also $-A^{c}_{n,m}(\bm{k})A^{b}_{m,n;a}(\bm{k})=A^{c}_{m,n}(\bm{k})A^{b}_{n,m;a}(\bm{k})$ under the BZ summation. The difference between the currents for right and left circular polarization, which is proportional to $\Im\sigma^{a;bc}_{\text{shift}}$ only, is therefore vanishing; and the shift current is associated with linearly polarized light. In the event of $\hat{\pazocal{T}}$ breaking, a circular shift current generally emerges as well as a non-stationary injection current \cite{Nagaosa2020,BSPVE}. We remark that the injection current can be equally computed with the method described in this work, but we do not show explicit results since the calculation is straightforward compared to that for the shift current \cite{AversaSipe,Sipe2000}, up to an extrinsic scattering rate that is often set phenomenologically.

\subsection{Berry connection and velocity in a local basis}
The single-particle crystalline eigenstates are generally expanded in a set of states \footnote{Which in practice is not an actual (complete) basis in this function space due to its finiteness. Nevertheless, we use this term referring to non-complete sets according to standard convention.} satisfying Bloch's theorem as
\begin{equation*}
\ket{\psi_{n,\bm{k}}}=\sum_{\mu}c_{\mu,n}(\bm{k})\ket{\phi_{\mu,\bm{k}}}
\end{equation*}
where $\mu$ is in principle a generic label and the coefficients are obtained from the generalized eigenvalue problem 
\begin{equation*}
\sum_{\mu'}\left[H_{\mu,\mu'}(\bm{k}) - E_{n}(\bm{k})S_{\mu,\mu'}(\bm{k})\right]c_{\mu',n}(\bm{k})=0
\end{equation*}
where the Hamiltonian $H_{\mu,\mu'}(\bm{k})$ and overlap $S_{\mu,\mu'}(\bm{k})$ matrix elements are the representations of the Hamiltonian $\hat{H}$ and identity $\hat{I}$ operator, respectively, in the set $\set{\ket{\phi_{\mu,\bm{k}}}}_{\mu}$ for each $\bm{k}$. In a local basis, which is repeated in each unit cell and labelled by the lattice vectors $\bm{R}$, the Bloch states can in turn be expanded in agreement with Bloch's theorem
\begin{equation*}
\ket{\phi_{\mu,\bm{k}}}=\frac{1}{\sqrt{N}}\sum_{\bm{R}}e^{i\bm{k}\bm{R}}\ket{\varphi_{\mu,\bm{R}}}
\end{equation*}
where $N=V/V_{\text{PUC}}\to\infty$ is the number of unit cells in the crystal. Therefore 
\begin{equation}\begin{aligned} \label{HkSk}
&H_{\mu,\mu'}(\bm{k})=\sum_{\bm{R}}e^{i\bm{k}\bm{R}}H_{\mu,\mu'}(\bm{R}), \text{ with }H_{\mu,\mu'}(\bm{R})=\bra{\varphi_{\mu,\bm{0}}}\hat{H}\ket{\varphi_{\mu',\bm{R}}}, \\
&S_{\mu,\mu'}(\bm{k})=\sum_{\bm{R}}e^{i\bm{k}\bm{R}}S_{\mu,\mu'}(\bm{R}), \text{ with }S_{\mu,\mu'}(\bm{R})=\bra{\varphi_{\mu,\bm{0}}}\ket{\varphi_{\mu',\bm{R}}}
\end{aligned}\end{equation}
due to the periodicity of $\hat{H}(\bm{r})$ and $\hat{I}$. 

The Berry connection matrix elements can then be expressed in the local basis by inserting the transformation $u_{n,\bm{k}}(\bm{r})=e^{-i\bm{k}\bm{r}}\sum_{\mu}c_{\mu,n}(\bm{k})\frac{1}{\sqrt{N}}\sum_{\bm{R}}e^{i\bm{k}\bm{R}}\varphi_{\mu,\bm{R}}(\bm{r})$, yielding after some algebra
\begin{equation*}
A^{b}_{m,n}(\bm{k})=
-i\left.\frac{\partial c^{*}_{\mu,m}(\bm{k}')}{\partial k^{\prime b}}\right\vert_{\bm{k}}S_{\mu,\mu'}(\bm{k})c_{\mu',n}(\bm{k}) + 
c^{*}_{\mu,m}(\bm{k})r^{b}_{\mu,\mu'}(\bm{k})c_{\mu',n}(\bm{k})
\end{equation*}
which can be compactly expressed in matrix form as
\begin{equation}\label{BerryConnection}
A^{b}(\bm{k})=-i\left.\frac{\partial c^{\dagger}(\bm{k}')}{\partial k^{\prime b}}\right\vert_{\bm{k}}S(\bm{k})c(\bm{k}) + c^{\dagger}(\bm{k})r^{b}(\bm{k})c(\bm{k})
\end{equation}
We have introduced the position matrix elements in the Bloch basis
\begin{equation}\label{positionME}
\bm{r}_{\mu,\mu'}(\bm{k})=\sum_{\bm{R}}e^{i\bm{k}\bm{R}}\bm{r}_{\mu,\mu'}(\bm{R}), \text{ with }\bm{r}_{\mu,\mu'}(\bm{R})=\bra{\varphi_{\mu,\bm{0}}}\hat{\bm{r}}\ket{\varphi_{\mu',\bm{R}}}=\int_{\mathbb{R}^{3}}\varphi_{\mu,\bm{0}}(\bm{r})^{*}\bm{r}\varphi_{\mu',\bm{R}}(\bm{r})d^{3}\bm{r}
\end{equation}
which is well defined in a localized basis, albeit the diagonal components depend on the origin choice. Indeed, a rigid shift of the form $\bm{r}\rightarrow\bm{r}+\bm{r}_{0}$ (which does not alter the position operator $\hat{\bm{r}}$ itself) with $\bm{r}_{0}$ restricted to the unit cell, results in $A^{b}_{m,n}(\bm{k})\rightarrow A^{b}_{m,n}(\bm{k})-\bm{r}_{0}\delta_{m,n}$. Nevertheless, it is easy to see that this arbitrary factor is cancelled in the GD $A^{b}_{m,n;a}$ rendering the shift conductivity invariant under this choice.

As it can be observed in \eqref{BerryConnection}, the calculation of both diagonal and off-diagonal Berry connections requires evaluating numerical derivatives with respect to $\bm{k}$. In order to avoid further complications in the term $\partial_{k^{a}}A^{b}_{n,m}$ of the GD $A^{b}_{n,m;a}$, such as the introduction of a second grid for derivatives or the increase in the required significant digits, we employ the identity 
\begin{equation}\label{BerryConnectionND} 
\bm{A}_{n,m}(\bm{k})=\frac{\bm{v}_{n,m}(\bm{k})}{i\omega_{n,m}(\bm{k})}\:,\;\;\;m\neq n
\end{equation}
which can be readily obtained by expanding $\bm{\nabla}_{\bm{k}}\left[\bra{u_{m,\bm{k}}}e^{-i\bm{k}\bm{r}}\hat{H}e^{i\bm{k}\bm{r}}\ket{u_{n,\bm{k}}}\right]=0$ for $m\neq n$. In \eqref{BerryConnectionND}, $\bm{v}_{n,m}(\bm{k})=\bra{\psi_{n,\bm{k}}}\bm{\hat{v}}\ket{\psi_{m,\bm{k}}}=\bm{v}_{m,n}(\bm{k})^{*}$ are the matrix elements (here in the set of eigenstates) of the velocity operator $\hat{\bm{v}}\equiv(i/\hbar)[\hat{H},\hat{\bm{r}}]$. It can be shown that the velocity matrix elements have the following form in the local basis \cite{esteve2023comprehensive} 
\begin{equation}\label{velocity}
\bm{v}_{n,m}(\bm{k})=\frac{1}{\hbar}\sum_{\mu,\mu'}  c^{*}_{\mu,n}(\bm{k})\left[
\bm{\nabla}_{\bm{k}}H_{\mu,\mu'}(\bm{k}) -E_{n}(\bm{k})\bm{\nabla}_{\bm{k}}S_{\mu,\mu'}(\bm{k})
+i\hbar\omega_{n,m}(\bm{k})\bm{r}_{\mu,\mu'}(\bm{k})
\right]c_{\mu',m}(\bm{k}) 
\end{equation}
which is independent on the origin choice. The $\bm{k}-$derivatives in this expression are all analytical, in particular $\bm{\nabla}_{\bm{k}}H_{\mu,\mu'}(\bm{k})=i\sum_{\bm{R}}e^{i\bm{k}\bm{R}}\bm{R}H_{\mu,\mu'}(\bm{R})$ and likewise for $S_{\mu,\mu'}(\bm{k})$. Therefore, $\partial_{k^{a}}A^{b}_{n,m}$ and $A^{b}_{n,m}$ can be computed employing \eqref{BerryConnectionND} with numerical derivatives only of the first and zeroth order, respectively. Inserting these into \eqref{SCgd1}, noting that the $f_{m,n}$ factor forces $m\neq n$ and that the derivatives of $\omega_{n,m}$ cancel out, one obtains
\begin{equation}\label{SCgd2} 
\begin{aligned}
&\sigma^{a;bc}_{\text{shift}}(\omega)=-\frac{ig_{s}\pi\abs{e}^3}{2\hbar^{2}V}\sum_{\bm{k}\in\text{BZ}}\sum_{m,n}
\frac{f_{m,n}}{\omega^{2}_{m,n}}\left[v^{b}_{m,n}v^{c}_{n,m;a}-v^{c}_{n,m}v^{b}_{m,n;a} \right]\delta(\omega_{m,n}-\omega)
=\\
&-\frac{ig_{s}\pi\abs{e}^3}{2\hbar^{2}V}\sum_{\bm{k}\in\text{BZ}}\sum_{m,n} 
\frac{f_{m,n}}{\omega^{2}_{m,n}}\left\{
v^{b}_{m,n}\left[ \frac{\partial v^{c}_{n,m}}{\partial k^{a}} - iv^{c}_{n,m}\left(A^{a}_{n,n}-A^{a}_{m,m}\right) \right]
-(b\leftrightarrow c)^{*} \right\} \delta(\omega_{m,n}-\omega)
\end{aligned}\end{equation}
where $(b\leftrightarrow c)^{*}$ represents the conjugated of the previous term inside the curly brackets with the components $b$ and $c$ permuted (even if $b=c$), and the $\bm{k}$ dependence has been omitted for brevity. Note that with $\hat{\pazocal{T}}$ symmetry the $-(b\leftrightarrow c)^{*}$ term is equivalent to $+(b\leftrightarrow c)$ under the BZ summation.

A subtle issue in \eqref{SCgd2} and other equivalent length gauge formulae is the evaluation of numerical derivatives of quantities that are not gauge-invariant, in particular the coefficients $c_{\mu,m}$ in the diagonal Berry connections, see \eqref{BerryConnection}, and the velocities $\bm{v}_{n,m}$ for $n\neq m$; which are respectively determined up to arbitrary phase factors $\theta_{m,\bm{k}}$ and $\theta_{m,n}^{\bm{k}}=\theta_{m,\bm{k}}-\theta_{n,\bm{k}}$. The continuity of \eqref{HkSk} in $\bm{k}$ makes all $c(\bm{k})$ (and $\bm{v}(\bm{k})$ by extension) also continuous, except in general for the phases. In order to fix a continuous gauge, we impose that 
\begin{equation*}
e^{i\theta_{m,\bm{k}}}\sum_{\mu}c_{\mu,m}(\bm{k})=\abs{\sum_{\mu}c_{\mu,m}(\bm{k})}\in\mathbb{R}\:,\;\;\forall m,\bm{k}
\end{equation*}
The necessary $\bm{k}-$ derivatives are well defined this way, and the $\theta_{m,\bm{k}}$ factors are all cancelled in the gauge-invariant \eqref{SCgd2} by a similar argument than in the time-reversal case above. The only remaining caveat is to keep track of the correct band indexing when a degeneracy occurs between the infinitesimally close $\bm{k}\pm\bm{h}$ points defining the numerical derivatives. However, this issue may be neglected by mapping the $\bm{k}-$summation to the interior of the irreducible Brillouin zone (IBZ), where only accidental degeneracies may occur, see Section \ref{SecIBZ}.

In 2D materials the out of plane tensor components, i.e., involving at least one index along the non-periodic direction $\bm{z}$, can be computed on equal footing (and likewise for 1D systems). Regarding the system as a periodic stacking of layers, $H_{\mu,\mu'}(\bm{R}_{z})$ \footnote{In charge-neutral 2D systems, the Coulomb potential may be replaced by the Parry potential instead of the usual Ewald potential in 3D \cite{doll2006analytical}.}, $S_{\mu,\mu'}(\bm{R}_{z})$, $\bm{r}_{\mu,\mu'}(\bm{R}_{z})$ are exponentially vanishing for inter-layer vectors $\bm{R}_{z}$ in the limit of large layer separation, making all $k^{z}-$derivatives null. In this case \eqref{BerryConnection} and \eqref{velocity} reduce to $A^{z}_{m,n}=-iv_{m,n}^{z}/\omega_{m,n}=\sum_{\mu,\mu'}c^{\dagger}_{m,\mu}r^{z}_{\mu,\mu'}c_{\mu',n}$. While the $a=z$ component in the shift conductivity tensor may not be of interest, in some point groups the IBZ summation requires the calculation of some of these components for $a=x,y$; in which case the numerical derivatives in \eqref{SCgd2} are cancelled and the expression is simplified significantly.   

An alternative treatment of \eqref{SCgd1} that is often found in the literature \cite{sturman1992photovoltaic,Sipe2000,Rappe2012} consists on the introduction of the shift vector, which involves the term $\bm{\partial_{k^{a}}}\Phi^{b}_{n,m}(\bm{k})$ where $v^{b}_{n,m}=\abs{v^{b}_{n,m}}e^{-i\Phi^{b}_{n,m}}$. This is obtained by noting that in the GD $\partial_{k^{a}}A^{b}_{n,m}=A^{b}_{n,m}\partial_{k^{a}}\log A^{b}_{n,m}$ when $A^{b}_{n,m}\neq0$. The term inside the curly brackets in \eqref{SCgd2}  is then equivalent to 
\begin{equation*}
-iv^{b}_{m,n}v^{c}_{n,m}\left[ \frac{\partial(\Phi^{b}_{n,m}+\Phi^{c}_{n,m})}{\partial k^{a}} + 2(A^{a}_{n,n}-A^{a}_{m,m}) +i\frac{\partial\log\abs{v^{c}_{n,m}/v^{b}_{n,m}}}{\partial k^{a}}  \right]
\end{equation*}
For linearly polarized light one can always rotate the coordinate system, initially based on the crystallographic structure, such that $\bm{E}(\omega)$ points along, say, the $b$ direction. Then only the $\sigma^{a;bb}$ component contributes to the current along $a$ in \eqref{current}, and the computation of the tensor is slightly simplified; in particular the modulus term vanishes in the last expression. This is, however, not advisable for practical calculations because it requires evaluating the tensor for each field direction with a different coordinate system, which may also hinder the obtention of the Hamiltonian matrix elements from the electronic structure code. In this work \eqref{SCgd2} is employed instead, since the computational cost is similar.

\subsection{Evaluation in Gaussian basis sets}
The evaluation of \eqref{SCgd2} from first-principles requires thus the knowledge of the matrix elements of $\hat{H}$, $\hat{I}$ and $\hat{\bm{r}}$ in the local basis for a sufficiently large number of lattice vectors 
The first one, $H_{\mu,\mu'}(\bm{R})$, must be evaluated self-consistently, typically in a DFT or hybrid DFT-HF (Hartree-Fock) scheme; and is generally expected to be provided by the corresponding electronic structure code for the chosen functional. The latter two, $S_{\mu,\mu'}(\bm{R})$ and $\bm{r}_{\mu,\mu'}(\bm{R})$ can be manually pre-computed from the (possibly optimized) atomic structure. If the local functions are harmonic Gaussian-type orbitals (GTOs), this can be done analytically. In that case $\mu$ is a multi-index labelling the atoms $a$ (located at $\bm{d}_{a}$) in the unit cell \footnote{Ghost atoms would be treated analogously as long as they preserve the space group symmetry.}, the pair of orbital quantum numbers $l,m$ ($m=-l,\dots,l$), the shells $\lambda$ discerning the harmonics with identical $l$ and, in the presence of spin dependent terms in $\hat{H}$ such as spin-orbit coupling (SOC) or magnetic ordering, the $m_{s}=\pm1/2$ spin quantum number. The contracted real GTOs $\varphi_{\mu,\bm{R}}:\mathbb{R}^{3}\to\mathbb{R}$ are then defined as \cite{dovesicrystal17,helgaker2013molecular}
\begin{equation*}
\varphi_{\mu,\bm{R}}(\bm{r})=N_{\lambda,l}\left[\sum_{j}c_{l,m,j}d_{\lambda,j}G(\alpha_{\lambda,j},\bm{r}-\bm{d}_{a}-\bm{R})\right] X_{l,m}(\bm{r}-\bm{d}_{a}-\bm{R})
\end{equation*}
where $N_{\lambda,l}$ and $c_{l,m,j}$ are normalization coefficients (see Appendix E of Ref. \cite{dovesicrystal17}), $d_{\lambda,j}$ and $\alpha_{\lambda,j}$ are the selected contraction coefficients and exponents, $G(\alpha,\bm{r})=e^{-\alpha r^{2}}$ are the Gaussian-type radial functions and $X_{l,m}(\bm{r})$ are the real solid harmonics. The latter are obtained from the (not normalized) spherical harmonics $Y_{l,m}(\bm{r})$ as 
\begin{equation*}
X_{l,m}(\bm{r})=\frac{r^{l}}{2}\cdot\left\{ \begin{aligned}
& Y_{l,\abs{m}}(\bm{r})+Y_{l,-\abs{m}}(\bm{r})\:,&\text{ if }m\geq0  \\
& -i\left[Y_{l,\abs{m}}(\bm{r})-Y_{l,-\abs{m}}(\bm{r})\right]\:,&\text{ if }m<0
\end{aligned} \right.
\end{equation*}
The central integrals that need be evaluated to obtain $S(\bm{k})$ and $\bm{r}(\bm{k})$ are then
\begin{equation}\label{integrals} \begin{aligned} 
&I_{3}(\bm{n},\bm{n}',\bm{r}_{0},\bm{r}'_{0},\alpha,\alpha')=I(n_{x},n'_{x},x_{0},x'_{0},\alpha,\alpha')\cdot I(n_{y},n'_{y},y_{0},y'_{0},\alpha,\alpha')\cdot I(n_{z},n'_{z},z_{0},z'_{0},\alpha,\alpha')\:,   \\
&I(n_{x},n'_{x},x_{0},x'_{0},\alpha,\alpha')\equiv\int_{\mathbb{R}}(x-x_{0})^{n_{x}}(x-x'_{0})^{n'_{x}}e^{-\alpha(x-x_{0})^{2}-\alpha'(x-x'_{0})^{2}}dx=I(n_{x},n'_{x},x_{0}-x'_{0},0,\alpha,\alpha')
\end{aligned}\end{equation}
The one-dimensional integrals appearing in $\bm{r}(\bm{k})$ are then computed as
\begin{equation*}
\int_{\mathbb{R}}(x-x_{0})^{n}(x-x'_{0})^{n'}xe^{-\alpha(x-x_{0})^{2}-\alpha'(x-x'_{0})^{2}}dx=x'_{0}I(n_{x},n'_{x},x_{0}-x'_{0},0,\alpha,\alpha') + I(n_{x},n'_{x}+1,x_{0}-x'_{0},0,\alpha,\alpha')
\end{equation*}
There are several methods to tabulate the one-dimensional integrals in \eqref{integrals}, e.g., by recursion over $n$ and $n'$. In this work we have instead employed the following master expression, which can be deduced from Ref. \cite{zwillinger2007table}
\begin{equation*}\begin{aligned}
&I(n,n',x_{0},0,\alpha,\alpha')= \\
&e^{-\tilde{\alpha}\:\alpha'x_{0}^{2}}\:x_{0}^{n+n'}\sqrt{\frac{\pi\tilde{\alpha}}{\alpha}}\:n!\sum_{k=0}^{n}(-1)^{n+k}\frac{(n'+k)!}{(n-k)!\:k!}\:\tilde{\alpha}^{n'+k}\sum_{h=0}^{\lfloor \frac{n'+k}{2} \rfloor}\frac{1}{(n'+k-2h)!\:h!}\left(\frac{1}{4\tilde{\alpha}\:\alpha\:x_{0}^{2}}\right)^{h}\:,\;\forall x_{0}\neq0
\end{aligned} \end{equation*}
where $\tilde{\alpha}\equiv\alpha/(\alpha+\alpha')$ and
\begin{equation*}
I(n,n',0,0,\alpha,\alpha')=\left\{\begin{aligned}
& \sqrt{\frac{\pi}{2^{n+n'}(\alpha+\alpha')^{n+n'+1}}}(n+n'-1)!! \:,&\text{ for }n+n'\text{ even}\\
& 0\:,&\text{ for }n+n'\text{ odd}
\end{aligned}\right.
\end{equation*}
We note that only one of $S_{\mu,\mu'}(\pm\bm{R})$ (and likewise for $\bm{r}_{\mu,\mu'}(\pm\bm{R})$) needs be computed for each $\bm{R}\neq\bm{0}$ since
\begin{equation*}
S_{\mu,\mu'}(-\bm{R})=S_{\mu',\mu}(\bm{R})\:,\;\;\bm{r}_{\mu,\mu'}(-\bm{R})=\bm{r}_{\mu',\mu}(\bm{R})-\bm{R}S_{\mu',\mu}(\bm{R})
\end{equation*}
and only the upper or lower triangle for $\bm{R}=\bm{0}$. If needed, the number of matrix elements may be further restricted such that only atoms in the asymmetric unit are considered in, say, the bra. The remaining entries can then be reconstructed by employing the (spinless) transformation properties of the real solid harmonics $\hat{g}^{-1}X_{l,m}(\bm{r})=X_{l,m}(g\bm{r})=\sum_{m'}\pazocal{D}^{l}_{m,m'}(g)X_{l,m'}(\bm{r})$, and the position operator $\hat{g}^{-1}\hat{r}^{b}\hat{g}=\sum_{b'}\pazocal{D}^{1}_{b,b'}(g)\hat{r}^{b'}$, where $g\in O(3)$ and $\pazocal{D}^{l}$ is the representation of $O(3)$ of angular momentum $l$ \cite{bir1974symmetry} ($l=1$ for $\bm{r}$). The result is 
\begin{equation}\label{AsymmUnit}\begin{aligned}
&S_{(\mathbbm{g}a,\lambda,l,m),(\mathbbm{g}'a',\lambda',l',m')}(\bm{R})=\sum_{m_{1},m_{1}'}\pazocal{D}^{l}_{m,m_{1}}(g)^{*}\pazocal{D}^{l'}_{m',m'_{1}}(g)S_{(a,\lambda,l,m_{1}),(\mathbbm{g}^{-1}\mathbbm{g}'a',\lambda',l',m_{1}')}(g^{-1}\bm{R}) \:, \\
&r^{b}_{(\mathbbm{g}a,\lambda,l,m),(\mathbbm{g}'a',\lambda',l',m')}(\bm{R})=\sum_{m_{1},m_{1}'}\pazocal{D}^{l}_{m,m_{1}}(g)^{*}\pazocal{D}^{l'}_{m',m'_{1}}(g)\sum_{b'}\cdot \\
&\left[ \pazocal{D}^{1}_{b,b'}(g)r^{b'}_{(a,\lambda,l,m_{1}),(\mathbbm{g}^{-1}\mathbbm{g}'a',\lambda',l',m_{1}')}(g^{-1}\bm{R}) + 
\pazocal{D}^{1}_{b,b'}(g^{-1})t^{b'}S_{(a,\lambda,l,m_{1}),(\mathbbm{g}^{-1}\mathbbm{g}'a',\lambda',l',m_{1}')}(g^{-1}\bm{R})
\right]
\end{aligned}\end{equation}
where $\mathbbm{g}=(g\vert\bm{t})$ is a general non-symmorphic operation in the crystallographic point group $F$ (or space group excluding lattice translations, $G/T$), and $\mathbbm{g}a$ represents the atom located at $\mathbbm{g}\bm{d}_{a}$. Note that the atom $\mathbbm{g}^{-1}\mathbbm{g}'a'$ may require a non-trivial lattice vector in order to be mapped to the unit cell, thus altering $g^{-1}\bm{R}$. If SOC is considered, then $\hat{g}^{-1}X_{l,m,m_{s}}(\bm{r})=\sum_{m',m'_{s}}\pazocal{D}^{l}_{m,m'}(g)\pazocal{D}^{1/2}_{m_{s},m'_{s}}(g^{-1})X_{l,m',m'_{s}}$, where $\pazocal{D}^{1/2}(c_{\theta})=\pazocal{D}^{1/2}(ic_{\theta})=e^{-i(\bm{\sigma}\bm{e})\theta/2}$ (in the $\set{\uparrow,\downarrow}$ basis order) is the projective representation of $O(3)$ of angular momentum $1/2$ which is even under inversion $i$; $\bm{\sigma}$ being the Pauli vector and $\bm{e}$ the counterclockwise rotation axis. The previous relations would be modified in consequence.

In plane wave schemes, $\hat{\bm{v}}$ is sometimes replaced by the momentum $\hat{\bm{p}}=-i\hbar\hat{\bm{\nabla}}$. This substitution is not exact in HF or hybrid DFT-HF schemes because $\hat{\bm{r}}$ and the Fock operator do not commute, or likewise when employing non-local pseudopotentials or including relativistic effect such as SOC. While the deviations in the final quantities are often not large \cite{Ivo2018}, the increase in computational cost in the Gaussian scheme is marginal enough to advise against the use of this approximation in general, except perhaps for extremely large unit cells. Regardless, the relevant integrals would be computed as
\begin{equation*}\begin{aligned}
&\int_{\mathbb{R}}(x-x_{0})^{n_{x}}e^{-\alpha(x-x_{0})^{2}}\frac{\partial}{\partial x}(x-x'_{0})^{n'_{x}}e^{-\alpha'(x-x'_{0})^{2}}dx = \\
&n'_{x}I(n_{x},n'_{x}-1,x_{0}-x'_{0},0,\alpha,\alpha')-2\alpha'I(n_{x},n'_{x}+1,x_{0}-x'_{0},0,\alpha,\alpha')
\end{aligned}\end{equation*}

\subsection{Velocity gauge}
Alternatively, the velocity gauge can be employed by imposing the minimal coupling $\hat{H}\left(-i\hbar\bm{\nabla}\right)\rightarrow\hat{H}\left(-i\hbar\bm{\nabla} + \abs{e}\bm{\pazocal{A}}\right)$, where $\bm{\pazocal{A}}$ is the vector potential. An analogous derivation in perturbation theory then yields \cite{von1981theory,BSPVE,Pedersen2017,Moore2019}
\begin{equation*}
\sigma^{a;bc}_{\text{total}}(\omega)=-\frac{g_{s}\abs{e}^{3}}{2\hbar^{2}\omega^{2}V}\sum_{\bm{k}\in\text{BZ}}\sum_{m,n,l}\left[
\frac{f_{m,n}v^{b}_{n,m}}{\omega_{m,n}-\omega+i\varepsilon}\left(\frac{v^{a}_{m,l}v^{c}_{l,n}}{\omega_{m,l}+i\varepsilon}-\frac{v^{c}_{m,l}v^{a}_{l,n}}{\omega_{l,n}+i\varepsilon}\right) - (b\leftrightarrow c)^{*}\right]
\end{equation*}
Both the shift and the injection conductivities are encoded in this formula. The former can be obtained by taking only the imaginary part of the product of complex denominators in the $\varepsilon\to0^{+}$ limit, which is equivalent to taking $\Re\sigma^{a;bc}_{\text{total}}$ under $\hat{\pazocal{T}}$ symmetry (i.e., considering linear polarization)
\begin{equation}\label{SCvg}
\sigma^{a;bc}_{\text{shift}}(\omega)=\frac{g_{s}\pi\abs{e}^3}{2\hbar^{2}V}\sum_{\bm{k}\in\text{BZ}}\sum_{m,n} 
\frac{f_{m,n}}{\omega^{2}_{m,n}} \Im\left[ 
v^{b}_{m,n}\sum_{l\neq m,n}\left( \frac{v^{a}_{n,l}v^{c}_{l,m}}{\omega_{n,l}} - \frac{v^{c}_{n,l}v^{a}_{l,m}}{\omega_{l,m}} \right)
 + (b\leftrightarrow c) \right]\delta(\omega_{m,n}-\omega)
\end{equation}
In contrast with \eqref{SCgd1} and \eqref{SCgd2}, the evaluation of \eqref{SCvg} avoids the numerical derivatives at the cost of a sum over all states $l$ that are external to the direct optical transition. Resulting from the completeness relation $\sum_{n}\ket{\psi_{n,\bm{k}}}\bra{\psi_{n,\bm{k}}}=\hat{I}$, in principle it must span the whole set of bands (which with localized bases is seldom demanding, computationally) even if they are not well represented above a certain window from the Fermi level, but the sum should nevertheless converge to the correct result when a sufficiently large basis is employed. While the length gauge explicitly involves only the pair of bands corresponding to the direct optical transition at the field frequency, a large basis should still be needed to properly reproduce the conduction bands involved. For grids of equal size, the evaluation of \eqref{SCvg} is more straightforward and less computationally demanding than \eqref{SCgd2}, however, the assumption of completeness (which is not strictly true in finite bases) makes the length gauge approach the most reliable in general.  

A similar expression to \eqref{SCvg} that is frequently employed in the literature 
is obtained within the length gauge by employing the following sum rule for the GD
\begin{equation}\label{SumRule}
A^{b}_{n,m;a}\rightarrow -\frac{1}{\omega_{n,m}}\left[
\left(A^{b}_{n,m}\Delta^{a}_{n,m}+A^{a}_{n,m}\Delta^{b}_{n,m}\right)+\sum_{l\neq m,n}\left(v^{b}_{n,l}A^{a}_{l,m}-A^{a}_{n,l}v^{b}_{l,m}\right) +iw^{a,b}_{n,m} \right]
\end{equation}
where $w^{a,b}_{n,m}\equiv\bra{\psi_{n,\bm{k}}}[\hat{r}^{a},\hat{v}^{b}]\ket{\psi_{m,\bm{k}}}=w^{b,a}_{n,m}$ and $\bm{\Delta}_{n,m}(\bm{k})\equiv\bm{v}_{n,n}(\bm{k})-\bm{v}_{m,m}(\bm{k})=\bm{\nabla}_{k}\omega_{n,m}(\bm{k})$ \footnote{The last identity results from the expansion of $\bm{\nabla}_{\bm{k}}[\bra{u_{n,\bm{k}}}e^{i\bm{k}\bm{r}}\hat{H}e^{i\bm{k}\bm{r}}\ket{u_{n,\bm{k}}}]=\bm{\nabla}_{\bm{k}}E_{n,\bm{k}}$}. This sum rule can be obtained by expanding $\partial_{k^{a}}\partial_{k^{b}}[\bra{u_{m,\bm{k}}}e^{-i\bm{k}\bm{r}}\hat{H}e^{i\bm{k}\bm{r}}\ket{u_{n,\bm{k}}}]=0$ and inserting the completeness relation above. Employing \eqref{SumRule} in \eqref{SCgd1} yields
\begin{equation}\begin{aligned}\label{SCts} 
&\sigma^{a;bc}_{\text{shift}}(\omega)\rightarrow -\frac{ig_{s}\pi\abs{e}^3}{2\hbar^{2}V}\sum_{\bm{k}\in\text{BZ}}\sum_{m,n} 
\frac{f_{m,n}}{\omega^{2}_{m,n}}\cdot \\ &\left\{ 
v^{b}_{m,n}\left[
\sum_{l\neq m,n}\left( \frac{v^{a}_{n,l}v^{c}_{l,m}}{\omega_{n,l}} - \frac{v^{c}_{n,l}v^{a}_{l,m}}{\omega_{l,m}} \right) -
\frac{v^{a}_{n,m}\Delta^{c}_{n,m}}{\omega_{n,m}} - iw^{a,c}_{n,m}
\right] - (b\leftrightarrow c)^{*} \right\}\delta(\omega_{m,n}-\omega)
\end{aligned}\end{equation}
where we note that the $\Delta^{a}$ terms cancel out. Clearly, this expression coincides with \eqref{SCvg} in the presence of $\hat{\pazocal{T}}$ symmetry, except for the last two terms inside the square brackets. The term $w^{a,c}$, which would clearly vanish in the absence of non-local terms in the Hamiltonian ($\hat{\bm{v}}=\hat{\bm{p}}$), is often computed in tight-binding or Wannier schemes by differentiating the Hamiltonian matrix \cite{cook2017design,Ivo2018} but its calculation in pure DFT is non-trivial and often ignored, leaving the two-band $\bm{\Delta}$ terms as the only difference in practice between the velocity gauge \eqref{SCvg} and length gauge with sume rule \eqref{SCts} expressions. While the contribution of these terms is usually small, in Section \ref{SecResults} we show that the proper agreement of the length gauge formula \eqref{SCgd1} or \eqref{SCgd2} is with the velocity gauge expression \eqref{SCvg}, at least if one neglects $w^{a,c}$.

\subsection{First-principles results and discussion}\label{SecResults}
\begin{figure}[H]
\centering 
\begin{subfigure}{0.48\textwidth}
    \includegraphics[width=\textwidth]{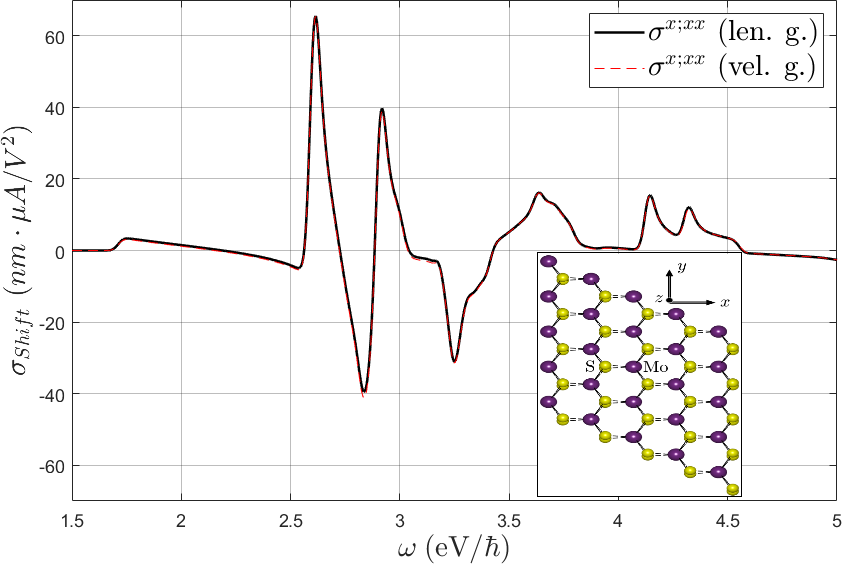}
    \caption{}
    \label{Fig1A}
\end{subfigure}
\begin{subfigure}{0.48\textwidth}
    \includegraphics[width=\textwidth]{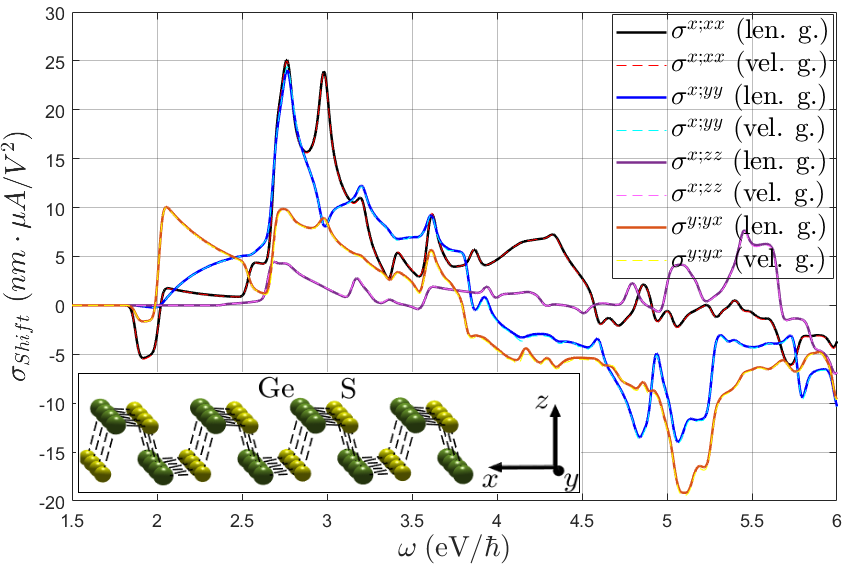}
    \caption{}
    \label{Fig1B}
\end{subfigure}
\begin{subfigure}{0.48\textwidth}
    \includegraphics[width=\textwidth]{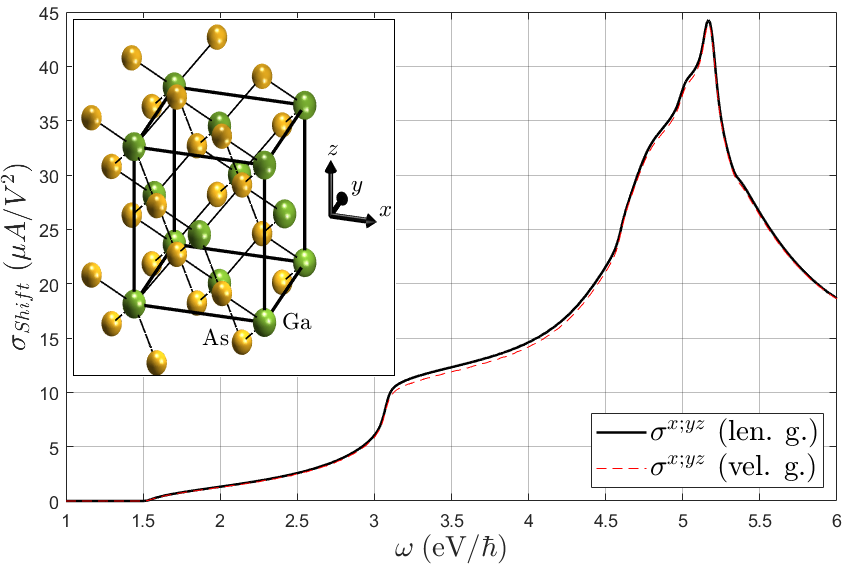}
    \caption{}
    \label{Fig1c}
\end{subfigure}
\begin{subfigure}{0.48\textwidth}
    \includegraphics[width=\textwidth]{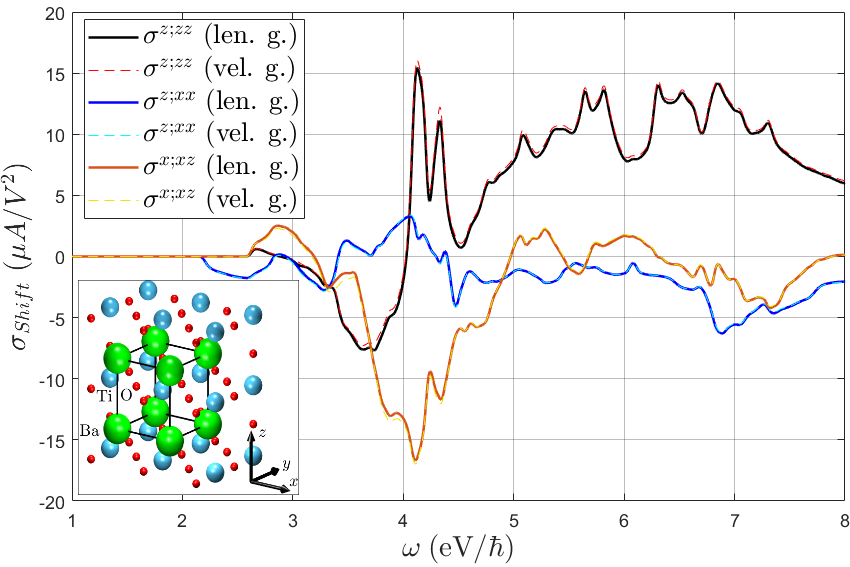}
    \caption{}
    \label{Fig1D}
\end{subfigure}
\caption{\label{Fig1} 
Shift conductivity tensor in (a) monolayer MoS$_{2}$, (b) monolayer GeS, (c) GaAs and (d) BaTiO$_{3}$ computed in the both the length \eqref{SCgd2} (solid lines) and velocity gauge \eqref{SCvg} (dashed lines). All linearly independent components are shown in each case, excluding $\sigma^{z;zx}$ in GeS for an out of plane current. IBZ summations with time-reversal symmetry \eqref{sumII} have been employed.}
\end{figure}

In Figure \ref{Fig1} we show the shift conductivity computed with large-sized Gaussian basis sets for some representative non-magnetic materials, in both the length \eqref{SCgd2} and velocity \eqref{SCvg} gauges. The self-consistent electronic structure problem has been solved with the \texttt{CRYSTAL} code \cite{crystal17,crystal23}, from which the $H_{\mu,\mu'}(\bm{R})$ are readily obtained. The input files for the self-consistent calculations can be found in the Supporting Information, in addition to the resulting band structures.

The starting points for the basis sets were the following: in MoS$_{2}$, def2-QZVP \cite{pritchard2019new} for S and pob-TZVP-rev2 \cite{laun2022bsse} for Mo; in GeS, def2-QZVP; in GaAs, m-pVDZ-PP-Heyd \cite{heyd2005energy}; in BaTiO$_{3}$, def2-QZVP with pseudo-potential (PP) from pob-TZVP-rev2 \cite{laun2021bsse} for Ba. In all cases the bases were modified in order to enable (or preserve) the convergence and obtain a sufficiently accurate band structure for the conduction bands in the energy ranges displayed in Figure \ref{Fig1}, except for GaAs which already presented a good dispersion with the unmodified Heyd basis. The standard GGA PBE functional \cite{perdew1996generalized} was used in MoS$_{2}$, GeS and BaTiO$_{3}$ in order to facilitate contrasting with the literature, while the short-range corrected hybrid HSE06 functional \cite{krukau2006influence} was employed in GaAs for the same reason. In the latter case, the use of a hybrid functional allows to obtain the experimental band gap at $\Gamma$ of $\sim1.5\:\text{eV}$ avoiding the use of a scissor correction or a $GW$ calculation.

The initial grids for the conductivity contained $2000\times2000$ points in the BZ for MoS$_{2}$, $1500\times1500$ for GeS, $400\times400\times400$ for GaAs and $200\times200\times200$ for BaTiO$_{3}$ \footnote{The grid choices were here influented by benchmarking purposes, and substantially coarser ones will often yield good results.}; and were subsequently restricted to the IBZ as explained in Section \ref{SecIBZ}, in particular employing \eqref{sumII} due to the absence of magnetism and the expressions from the list for the corresponding space groups: 187 ($D_{3h}$) reduced to 2D for MoS$_{2}$, 31 ($C_{2v}$) reduced to 2D with $c_{2,x}$ contained in the lattice plane for GeS, 216 ($T_{d}$) for GaAs and 99 ($C_{4v}$) for BaTiO$_{3}$. The delta function in the $\sigma_{\text{shift}}^{a;bc}$ expressions has been approximated by a narrow normal distribution $\delta(x)\sim\frac{1}{\sqrt{2\pi}\sigma}\exp\left[-\frac{x^{2}}{2\sigma^{2}}\right]$ with standard deviation $\sigma=20\:\text{meV}$ in all cases. The absence of (unphysical) rapid fluctuations in the curves indicates that this value is not small for the chosen $\bm{k}-$grids. The numerical derivatives in the length gauge expression \eqref{SCgd2} have been symmetrized as $\partial_{k^{a}}f(\bm{k})\sim\frac{1}{2h}( f(\bm{k}+h\bm{e}^{a}) - f(\bm{k}-h\bm{e}^{a}) )$, with $h=\frac{10^{-7}}{a_{0}}$ in all cases. The direct lattice summations in \eqref{HkSk} and \eqref{positionME} have been truncated to the first (by length) 179 vectors in MoS$_{2}$, 120 in GeS, 179 in GaAs and 260 in BaTiO$_{3}$. 

The agreement with the results in the literature is generally good \cite{BSPVE,rangel2017large,wang2017first,Ivo2018,Rappe2012,gjerding2021recent}, specially taking into account that moderate discrepancies can be found commonly due to the high sensibility of $\sigma_{\text{shift}}$ to the lattice parameters, atomic coordinates and electronic eigenfunctions \cite{cook2017design,schankler2021large,zhang2022tailoring}, in conjunction with the different calculation methods (as outlined in the previous sections) and convergence parameters such as the BZ grid or the broadening of the delta functions. The proper description of the eigenstates in the $\hbar\omega$ energy range is critical to obtain satisfactory results, hence in principle the largest possible basis set allowed by convergence should be used, typically a reduced QZVP or augmented TZVP. Nevertheless, in materials with not particularly delocalized empty conduction states a smaller, but properly calibrated basis could suffice. It should be kept in mind that the truncation of the $\bm{R}-$sums may need to be loosed if smaller Gaussian exponents are introduced. 

It can be observed in Figure \ref{Fig1} that the results of both length and velocity gauges are virtually identical in all cases, even if the external sum in \eqref{SCvg} is truncated in practice by the finiteness of the basis. This is in agreement with the results of Reference \cite{passos2018nonlinear} for third-order calculations in graphene. However, the inclusion of the two-band $\bm{\Delta}$ terms in \eqref{SCts} by the sum rule induces a small, but noticeable discrepancy in some of the $\sigma^{a;bc}_{\text{shift}}$ components, specifically $\sigma^{x;yy}$, $\sigma^{y;yx}$ in GeS and $\sigma^{z;xx}$ in BaTiO$_{3}$, while all other components (including MoS$_{2}$ and GaAs) were not visibly affected. Most of the unaltered components have a symmetry reason to remain so: the $\bm{\Delta}$ terms clearly do not contribute to $a=b=c$ components in \eqref{SCts}, nor to $b=c=z$ components in 2D materials since $\Delta^{z}=0$. Furthermore, it can also be seen from \eqref{SCts} that a sum over even permutations of $(a,b,c)$ cancels the $\bm{\Delta}$ terms, which is precisely underlying in GaAs as can be seen from the folded summation formula for $T_{d}$ in Section \ref{SecIBZ}. Only the $\sigma^{x;xz}$ component in BaTiO$_{3}$ cannot be explained by symmetry reasons, albeit in this case only a term of the form $v^{z}_{n,m}v^{x}_{m,n}\Delta^{x}_{n,m}$ contributes under $\hat{\pazocal{T}}$ symmetry, which is most likely small for numerical reasons in the dispersion along $k^{x}$.

\begin{figure}[H]
\centering 
\begin{subfigure}{0.48\textwidth}
    \includegraphics[width=\textwidth]{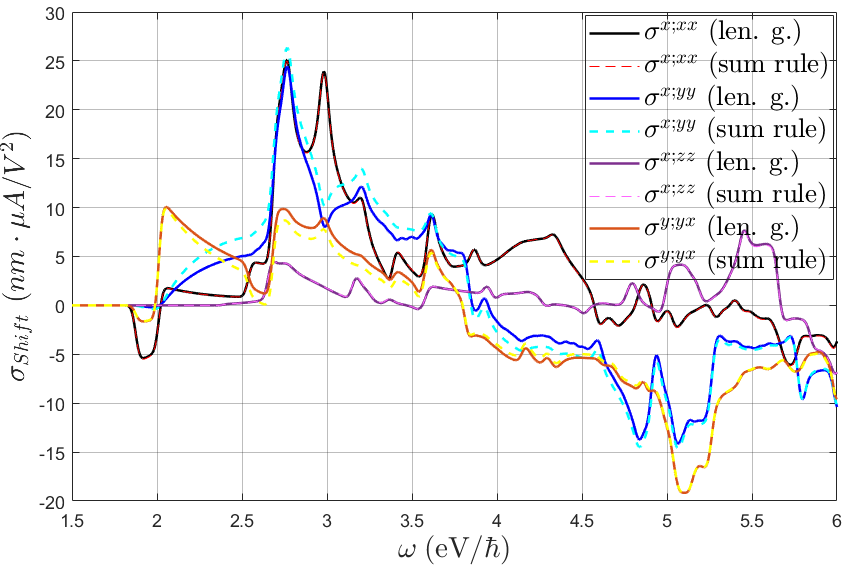}
    \caption{}
    \label{Fig2A}
\end{subfigure}
\begin{subfigure}{0.48\textwidth}
    \includegraphics[width=\textwidth]{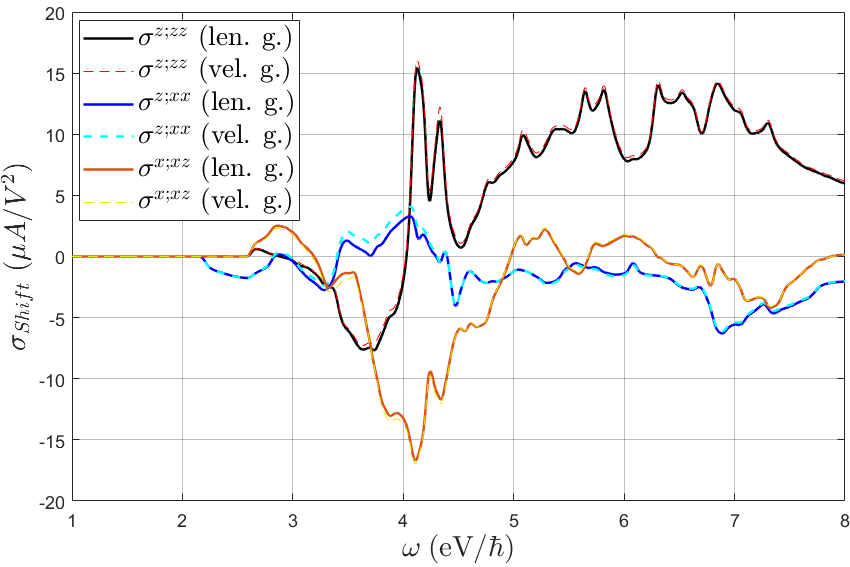}
    \caption{}
    \label{Fig2B}
\end{subfigure}
\caption{\label{Fig2} 
Shift conductivity tensor in (a) monolayer GeS and (b) BaTiO$_{3}$ computed in the length gauge \eqref{SCgd2} (solid lines) and employing the sum rule \eqref{SCts} (dashed lines) excluding the $w^{a,c}$ term. IBZ summations with time-reversal symmetry \eqref{sumII} have been employed.}
\end{figure}

\section{Irreducible Brillouin Zone summation} \label{SecIBZ}
The proper use of GTOs ensures that the crystalline eigenstates indeed transform according to the irreducible representations of the space group $G$, and $\hat{\mathbbm{g}}\psi_{n,\bm{k}}(\bm{r})=\psi_{n,g\bm{k}}(\bm{r}-\bm{t}-\bm{R})=\psi_{n,\bm{k}}(\mathbbm{g}^{-1}\bm{r})$, $\forall\mathbbm{g}=(g\vert\bm{t}+\bm{R})\in G$. Then, since the velocity operator transforms under the corresponding change of coordinates as $\hat{\mathbbm{g}}^{-1}\hat{v}^{b}\hat{\mathbbm{g}}=\sum_{b'}\pazocal{D}^{1}_{b,b'}(g)\hat{v}^{b'}$ \footnote{Note that the translations commute with $\hat{H}$ as fixed parameters} (see discussion around \eqref{AsymmUnit} for context) and likewise for $\hat{\partial}_{k^{b}}$, the following identities must satisfied, up to an arbitrary phase factor in the eigenstates that has no impact in the conductivity
\begin{equation}\label{vunit}
v^{b}_{m,n}(g\bm{k})=\sum_{b,b'}\pazocal{D}^{1}_{b,b'}(g)v^{b'}_{m,n}(\bm{k})\;,\;
\partial_{(gk^{b})}=\sum_{b,b'}\pazocal{D}^{1}_{b,b'}(g)\partial_{k^{b'}}\;,\;
A^{b}_{m,n}(g\bm{k})=\sum_{b,b'}\pazocal{D}^{1}_{b,b'}(g)A^{b'}_{m,n}(\bm{k})\;,\;\forall g\in F
\end{equation}
where $F$ is the (isogonal) point group of the material, formed by disregarding all translations of the space group. On the other hand, under the time-reversal operation $\hat{\pazocal{T}}\hat{v}^{b}\hat{\pazocal{T}}^{-1}=-\hat{v}^{b}$ (odd power of time units), so that recalling from Section \ref{secLG} the action of $\hat{\pazocal{T}}$ on the Berry connection, if time-reversal is a symmetry then
\begin{equation}\label{vantiunit}
v^{b}_{m,n}(-\bm{k})=-v^{b}_{n,m}(\bm{k})\;,\;
A^{b}_{m,n}(-\bm{k})=A^{b}_{n,m}(\bm{k})
\end{equation}
where again the arbitrary phase factors have been already cancelled. We note that \eqref{vunit}, \eqref{vantiunit} hold irrespective of whether SOC is included, since the space group is the same \footnote{The space group can be regarded as unaltered if projective representations of the little groups are employed \cite{bir1974symmetry}, which do not appear in this derivation regardless. Alternatively, vector (standard) representations of enlarged little groups can be considered \cite{bradley2010mathematical}. On the other hand, we disregard the extra spin-only operations of the spin-space group that may appear without SOC. These may possibly introduce further restrictions, but would not overrule the present ones in any case.} and the representations of the eigenstates are not being used (but rather of the operators), and the sign of $\hat{\pazocal{T}}^{2}=\pm\hat{I}$ has no impact. Consequently, the following analysis is also valid irrespective of SOC.

These transformation properties can now be exploited to reduce the summations over the BZ in \eqref{SCgd1}, \eqref{SCgd2} or \eqref{SCts} to properly weighted sums over the IBZ or representation domain, from which the whole BZ is reconstructed by applying point group symmetries (possibly including $\pazocal{T}$). This ensures that each sampled $\bm{k}$ point provides unique information, reducing the computation time of the original BZ grid by approximately the order the of the point group, and avoids the touching of symmetry-enforced degeneracies which are troublesome for computing the numerical derivatives in the length gauge. Furthermore, the identification of the finite tensor entries and the linear dependencies between them is straightforward from the IBZ expressions. 

In order to avoid exceptions with duplicated points, we hereafter assume that the BZ sampling does not contain any high symmetry point (or, to be precise, any point whose little group is not trivial), which is without loss of generality in sufficiently fine grids. Several cases are distinguished depending on time-reversal symmetry, which determines the magnetic point group $M$.
\begin{itemize}
\item Magnetic point group of type I: $M$ contains only unitary operations, i.e., $M=F$ and $\pazocal{T}$ is excluded. Then by \eqref{vunit} the first term in \eqref{SCgd2} satisfies
\begin{equation*}\begin{aligned}
&\sum_{\bm{k}\in\text{BZ}}v^{b}_{m,n}(\bm{k})\left.\frac{\partial v^{c}_{n,m}(\bm{k}')}{\partial k^{\prime a}}\right\vert_{\bm{k}}=\sum_{g\in F}\sum_{\bm{k}\in\text{IBZ}}v^{b}_{m,n}(g\bm{k})\left.\frac{\partial v^{c}_{n,m}(\bm{k}')}{\partial k^{\prime a}}\right\vert_{g\bm{k}}=\\
&\sum_{\bm{k}\in\text{IBZ}}\sum_{a',b',c'}\mathbb{D}^{F}_{a,a';b,b';c,c'}v^{b'}_{m,n}(\bm{k})\left.\frac{\partial v^{c'}_{n,m}(\bm{k}')}{\partial k^{\prime a'}}\right\vert_{\bm{k}}
\end{aligned}\end{equation*}
and likewise for the other terms, also in the form of \eqref{SCgd1} or \eqref{SCts}. Here we have defined, for any subset $L\in O(3)$,
\begin{equation}\label{D1D1D1}
\mathbb{D}^{L}_{a,a';b,b';c,c'}\equiv\sum_{g\in L}\pazocal{D}^{1}_{a,a'}(g)\pazocal{D}^{1}_{b,b'}(g)\pazocal{D}^{1}_{c,c'}(g)
\end{equation}
Then, writing the $\bm{k}-$integrand explicitly for any of the forms \eqref{SCgd1}, \eqref{SCgd2} or \eqref{SCts}, 
\begin{equation}\label{sigmaK}
\sigma^{a;bc}_{\text{shift}}\equiv -\frac{i}{N_{\bm{k}}}\sum_{\bm{k}\in\text{BZ}}\left[\tilde{\sigma}^{a;bc}_{\text{shift}}(\bm{k})-\tilde{\sigma}^{a;cb}_{\text{shift}}(\bm{k})^{*}\right]
\end{equation}
it follows that 
\begin{equation}\label{sumI}
\sigma^{a;bc}_{\text{shift}}=-\frac{i}{\abs{F}N_{\bm{k}}^{\text{IBZ}}}\sum_{\bm{k}\in\text{IBZ}}\sum_{a',b',c'}\mathbb{D}^{F}_{a,a';b,b';c,c'}\left[\tilde{\sigma}^{a';b'c'}_{\text{shift}}(\bm{k})-\tilde{\sigma}^{a';c'b'}_{\text{shift}}(\bm{k})^{*}\right]
\end{equation}
where we have used that $N_{\bm{k}}=\abs{F}N_{\bm{k}}^{\text{IBZ}}$, with $\abs{F}$ the order of $F$ and $N_{\bm{k}}^{\text{IBZ}}$ the number of grid points in the IBZ. It is easy to see that $\mathbb{D}^{F}_{a,a';b,b';c,c'}=0$ for any group with inversion symmetry $i\in F$, hence $\sigma^{a;bc}_{\text{shift}}=0$ in agreement with the obvious requirement from perturbation theory.

\item Magnetic point group of type II: time-reversal is a symmetry by itself, i.e.,  $M=F+\pazocal{T}F$. This precludes any permanent magnetic ordering. Then by \eqref{vunit} and \eqref{vantiunit} 
\begin{equation*}\begin{aligned}
&\sum_{\bm{k}\in\text{BZ}}v^{b}_{m,n}(\bm{k})\left.\frac{\partial v^{c}_{n,m}(\bm{k}')}{\partial k^{\prime a}}\right\vert_{\bm{k}}=\frac{1}{g_{i}}\sum_{g\in F}\sum_{\bm{k}\in\text{IBZ}}\left[v^{b}_{m,n}(g\bm{k})\left.\frac{\partial v^{c}_{n,m}(\bm{k}')}{\partial k^{\prime a}}\right\vert_{g\bm{k}} + 
v^{b}_{m,n}(-g\bm{k})\left.\frac{\partial v^{c}_{n,m}(\bm{k}')}{\partial k^{\prime a}}\right\vert_{-g\bm{k}}\right]=\\
&\frac{1}{g_{i}}\sum_{\bm{k}\in\text{IBZ}}\sum_{a',b',c'}\mathbb{D}^{F}_{a,a';b,b';c,c'}2i\Im\left[v^{b'}_{m,n}(\bm{k})\left.\frac{\partial v^{c'}_{n,m}(\bm{k}')}{\partial k^{\prime a'}}\right\vert_{\bm{k}}\:\right]
\end{aligned}\end{equation*}
where $g_{i}=2$ if $i\in F$ or in the 2D cases with out of plane $c_{2,z}$ rotational symmetry, and $g_{i}=1$ otherwise. The group $F$ is again forced to be non-centrosymmetric, but time-reversal symmetry effectively halves the IBZ with respect to the type I case by introducing a relation between the $\pm\bm{k}$ pairs (except in the cases where $g_{i}=2$). The general shift current tensor \eqref{sigmaK} thus satisfies
\begin{equation}\label{sumII}
\sigma^{a;bc}_{\text{shift}}=\frac{1}{\abs{F}N^{\text{IBZ}}_{\bm{k}}}\sum_{\bm{k}\in\text{IBZ}}\sum_{a',b',c'}\mathbb{D}^{F}_{a,a';b,b';c,c'}\Im\left[\tilde{\sigma}^{a';b'c'}_{\text{shift}}(\bm{k})+\tilde{\sigma}^{a';c'b'}_{\text{shift}}(\bm{k})\right]
\end{equation} 
where $N_{\bm{k}}=2\abs{F}N_{\bm{k}}^{\text{IBZ}}/g_{i}$. In agreement with Section \ref{secLG}, $\sigma^{a;bc}_{\text{shift}}$ is real in this case. 

\item Magnetic point group of type III: exactly half of the unitary operations are paired with time-reversal, i.e., $M=H+\pazocal{T}(F-H)$ where $H\subset F$ is an ordinary point group and $F-H$ is thus not a group. In this case \eqref{vunit} is valid for $g\in H$ whereas \eqref{vantiunit} must be used in combination with \eqref{vunit} for $g\in F-H$, and the IBZ is defined by $F$ as in type I. Hence, symmetrizing in $\pm\bm{k}$,
\begin{equation*}\begin{aligned}
&\sum_{\bm{k}\in\text{BZ}}v^{b}_{m,n}(\bm{k})\left.\frac{\partial v^{c}_{n,m}(\bm{k}')}{\partial k^{\prime a}}\right\vert_{\bm{k}}=\frac{1}{2}
\sum_{\bm{k}\in\text{IBZ}}\sum_{\bm{k}'=\pm\bm{k}}\left[\sum_{g\in H}v^{b}_{m,n}(g\bm{k}')\left.\frac{\partial v^{c}_{n,m}(\bm{k}'')}{\partial k^{\prime\prime a}}\right\vert_{g\bm{k}'} + \right. \\
&\left. \sum_{\tilde{g}\in F-H}v^{b}_{m,n}(-\tilde{g}\bm{k}')\left.\frac{\partial v^{c}_{n,m}(\bm{k}'')}{\partial k^{\prime\prime a}}\right\vert_{-\tilde{g}\bm{k}'}  \right]= \\
&
\sum_{\bm{k}\in\text{IBZ}}\sum_{a',b',c'}\left[ \mathbb{D}^{H}_{a,a';b,b';c,c'}v^{b'}_{m,n}(\bm{k})\left.\frac{\partial v^{c'}_{n,m}(\bm{k}')}{\partial k^{\prime a'}}\right\vert_{\bm{k}} - \mathbb{D}^{F-H}_{a,a';b,b';c,c'}v^{b'}_{n,m}(\bm{k})\left.\frac{\partial v^{c'}_{m,n}(\bm{k}')}{\partial k^{\prime a'}}\right\vert_{\bm{k}}\:  \right] 
\end{aligned}\end{equation*}
and, noting that $\mathbb{D}^{F-H}=\mathbb{D}^{F}-\mathbb{D}^{H}$, \eqref{sigmaK} can be expressed as
\begin{equation}\label{sumIII} \begin{aligned}
&\sigma^{a;bc}_{\text{shift}}=-\frac{i}{\abs{F}N^{\text{IBZ}}_{\bm{k}}}\sum_{\bm{k}\in\text{IBZ}}\sum_{a',b',c'} \\
&\left[ 2\mathbb{D}^{H}_{a,a';b,b';c,c'}\Re\left(\tilde{\sigma}^{a';b'c'}_{\text{shift}}(\bm{k})-\tilde{\sigma}^{a';c'b'}_{\text{shift}}(\bm{k})\right) - 
\mathbb{D}^{F}_{a,a';b,b';c,c'}\left(\tilde{\sigma}^{a';b'c'}_{\text{shift}}(\bm{k})^{*}-\tilde{\sigma}^{a';c'b'}_{\text{shift}}(\bm{k})\right)      \right]
\end{aligned}\end{equation}
If the crystalline structure is centrosymmetric disregarding magnetism, there are three cases upon magnetization. The first one is that $i$ is completely removed by the magnetization, $i\notin F$ and $M$ is of type I or III, thus it imposes no restrictions on $\sigma^{a;bc}_{\text{shift}}$. The second one is $i\in H\subset F$, which implies that $\mathbb{D}^{H}=\mathbb{D}^{F}=0$ and $\sigma^{a;bc}_{\text{shift}}=0$. The third one is $i\in F-H$, which implies that $\mathbb{D}^{F}=0$ and $\Re\sigma^{a;bc}_{\text{shift}}=0$. Therefore, a centrosymmetric material may host a finite shift current upon magnetization as long as inversion is not a symmetry by itself but in combination with time-reversal (often termed $\pazocal{P}\pazocal{T}$ symmetry), a situation that is most common in antiferromagnets and which by \eqref{current} and \eqref{sumIII} yields a shift current under circular (or elliptical) polarization.
\end{itemize}

Note that \eqref{sumI}, \eqref{sumII}, \eqref{sumIII} also include the case of 2D materials with the global symmetry $\sigma_{h}\in F$ (mirror plane parallel to the lattice), since the required $1/2$ factor in the $\sum_{\bm{k}\in\text{BZ}}\to(1/2)\sum_{g\in F}\sum_{\bm{k}\in\text{IBZ}}$ substitution is absorbed in $\abs{F/C_{1h}}=N_{\bm{k}}/N_{\bm{k}}^{\text{IBZ}}$.

Therefore, for a given material one must evaluate \eqref{sumI}, \eqref{sumII} or \eqref{sumIII} according to its magnetic point group, which requires computing \eqref{D1D1D1} for the appropriate subgroup of $O(3)$, and parametrizing the IBZ according to its space group since the lattice type determines the BZ. In the list at the end of this section we show, for each (unitary) non-centrosymmetric point group, the values of \eqref{D1D1D1} in the simplified notation $aa';bb';cc'$, excluding the vanishing components as well as those that are redundant due to the permutation properties $aa';bb';cc'=aa';cc';bb'=bb';aa';cc'=bb';cc';aa'=cc';aa';bb'=cc';bb';aa'$ and the reciprocity relation $a'a;b'b;c'c=aa';bb';cc'$. Afterwards the finite components of the $\sigma^{a;bc}_{\text{shift}}$ tensor and the relations between them are displayed, omitting the persistent $\sigma^{a;bc}_{\text{shift}}=(\sigma^{a;cb}_{\text{shift}})^{*}$ and specifying which of \eqref{sumI} or \eqref{sumII} are finite if only one of them is \footnote{For example, $\Im\sigma^{x;yz}_{I}$ indicates that \eqref{sumI} is purely imaginary and \eqref{sumII} is null for the $a=x,b=y,c=z$ component. Note that $\sigma^{a;bc}_{I}=\sigma^{a;bc}_{II}$ implies that \eqref{sumI} is real. We omit the mandatory equalities between \eqref{sumI} and \eqref{sumII} for $b=c$ components.}. When multiple orientations along the Cartesian axes are possible, we indicate the orientation of the generating operations with the notation $c_{n,a}$ for a counterclockwise $2\pi/n-$fold rotation with axis along $a$, $\sigma_{a}$ for a reflection through a plane perpendicular to $a$ and $s_{n,a}$ for the improper rotation $c_{n,a}\sigma_{a}$. If a different orientation is employed, then there exists a transformation $r\in O(3)$ of our coordinate system that matches the chosen orientation, and the new conductivity tensor $\overline{\sigma}^{a;bc}$ can be computed from the one given here as $\overline{\sigma}^{a;bc}=\sum_{a',b',c'}\sigma^{a';b',c'}$, which will often result in a simple permutation of coordinates.

On the other hand, the grid of the IBZ can be obtained (non-uniquely) as a restriction of a larger grid by imposing a set of linear constraints on the coefficients $\alpha_{i}$ of the $\bm{k}=\sum_{i=1}^{\text{dim}}\alpha_{i}\bm{G}_{i}$ points (``dim'' being the dimension of the lattice) in the basis of reciprocal lattice vectors $\bm{G}_{i}$. Specifically, 
\begin{equation}\label{IBZeq}
\text{IBZ}=\hat{R}\set{\sum_{i=1}^{\text{dim}}\alpha_{i}\bm{G}_{i}\:\vert\: UA^{t}\bm{\alpha}<u}
\end{equation}
where $U$ and $u$ depend on the space group and are given in the list below for a particular choice of reciprocal lattice vectors $\bm{G}^{\text{ref}}_{i}$, which are those of Reference \cite{bradley2010mathematical} (see tables 3.1 and 3.3 therein) and which we indicate in the list in $2\pi$ units for each lattice type under the first point group for which they appear. The coefficients $\alpha_{i}$ should initially be spanning the $(-1,1)$ interval because, while a range of length 1 for each would suffice, it is not guaranteed that our specific parametrization yields the same range for each coefficient \footnote{Note that in this case a $N_{1}\times N_{2}\times N_{3}$ grid for the $\alpha_{i}$ coefficients corresponds to a $(N_{1}/2)\times(N_{2}/2)\times(N_{3}/2)$ grid of the BZ}. In \eqref{IBZeq}, $A$ and $R$ are introduced to allow for different sets of lattice vectors: $A$ is the transformation matrix relating both sets of reciprocal vectors as
\begin{equation*}
\begin{pmatrix} \bm{G}_{1}\\ \vdots \\ \bm{G}_{\text{dim}} \end{pmatrix} =
A \begin{pmatrix} \bm{G}^{\text{ref}}_{1}\\ \vdots \\ \bm{G}^{\text{ref}}_{\text{dim}} \end{pmatrix}
\end{equation*}
and $\hat{R}$ is the operator represented by the $R\in SO(\text{dim})$ rotation matrix that relates both BZs, acting on the set of $\bm{k}$ points to its right. The parametrized IBZs lie within the BZ except for the triclinic and monoclinic systems. The space groups $G$ are labelled by their international number, and the different variations of IBZs that cannot be obtained through rotations are considered. 

The list can also be employed with 2D materials in the following way: first identify the 3D space group that is compatible with the two-dimensional structure, taking into account that in this case only groups with $\bm{G}^{\text{ref}}_{3}\parallel z$ and lacking non-symmorphic translations along $z$ can be compatible. Then, if $F=C_{2}$ or $C_{2v}$, permute two columns of $U$ to match the $c_{2}$ axis orientation \footnote{For example, in the GeS $C_{2v}$ ($c_{2x}$) geometry of the present work, the first and third columns are permuted.}, if $F=C_{s}$ ($\sigma_{y}$) do the columns permutation (123)$\to$(132), if $F=C_{s}$ ($\sigma_{x}$) do the columns permutation (123)$\to$(231); and if any of these 3 permutations was performed, do the same permutation of the $(x,y,z)$ indices in $aa';bb';cc'$ and $\sigma^{a;bc}$. Finally, eliminate all rows in $U$ and $u$ whose first two entries in $U$ are zero. The reduction of IBZs with $\hat{\pazocal{T}}$ has been chosen such that the last step yields the correct IBZ in 2D (invariant with $c_{2,z}$ or halved without it) with all compatible space groups.   

Either the red or the blue rows in $U$ and $u$ are included for a given system, in particular the red rows (excluding the blue) are included when the IBZ is not affected by $\pazocal{T}$, and the blue rows (excluding the red) when the IBZ is halved by $\pazocal{T}$. This can be determined for the 4 types of magnetic space groups \cite{bradley2010mathematical}, which we denote as $MG$, as an application of the previous discussion of magnetic point groups, in addition to the specific formula for $\sigma^{a;bc}_{\text{shift}}$. 
\begin{itemize}
\item Magnetic space group of type I: $MG=G$ $\rightarrow$ red rows, equation \eqref{sumI}.
\item Magnetic space group of type II: $MG=G+\pazocal{T}G$ $\rightarrow$ blue rows, equation \eqref{sumII}.
\item Magnetic space group of type III: $MG=\tilde{G}+\pazocal{T}(G-\tilde{G})$, where $\tilde{G}$ is an space group whose point group has half the order of $G$'s point group, the latter of which determines the IBZ $\rightarrow$ red rows, equation \eqref{sumIII}.
\item Magnetic space group of type IV: $MG=G+\pazocal{T}(e\vert\bm{t}_{0})G$, where $(e\vert\bm{t}_{0})\notin G$ is a pure translation. The IBZ is defined by $G$, for our purpose with $\pazocal{T}$ since the extra translation, as in non-symmorphic groups, is inconsequential $\rightarrow$ blue rows, equation \eqref{sumII}.  
\end{itemize}

\vspace{1cm}
\begin{itemize}
      \hrule\vspace{0.05cm}\hrule
      \vspace{-0.2cm}
\item $F=C_{1}$
\begin{equation*}
aa';bb';cc'=\delta_{a,a'}\delta_{b,b'}\delta_{c,c'}
\end{equation*}
\begin{equation*}
\text{No restrictions on } \sigma^{a;bc}
\end{equation*}
\begin{itemize}
\item $G=1$ $[\Gamma_{t}]$
\begin{equation*}
\bm{G}_{i} \text{ arbitrary, }(A=R=I),\; U=\begin{pmatrix}-1&0&0\\ \redrow{1}{0}{0}\\ \bluerow{1}{0}{0} \\0&-1&0\\ 0&1&0\\ 0&0&-1\\ 0&0&1 \end{pmatrix},\; u=\begin{pmatrix} 0\\ \textcolor{red}{1}\\ \textcolor{blue}{1/2}\\ 0\\1\\0\\1 \end{pmatrix}
\end{equation*}
\end{itemize}

\item $F=C_{2}$ ($c_{2,z}$)
\begin{equation*}
zz;zz;zz=2,\; zz;xx;xx=2,\; zz;yy;yy=2,\; xx;yy;zz=2
\end{equation*}
\begin{equation*}
\sigma^{z;zz},\; \sigma^{z;xx},\; \sigma^{z;yy},\; \sigma^{x;xz},\; \sigma^{y;yz},\; \sigma^{x;yz},\; \sigma^{y;zx},\; \sigma^{z;xy}
\end{equation*}
\begin{itemize}
\item $G=3-4$ [$\Gamma_{m}$]
\begin{equation*}
\begin{pmatrix}\bm{G}_{1}^{\text{ref}}\\\bm{G}_{2}^{\text{ref}}\\\bm{G}_{3}^{\text{ref}} \end{pmatrix}=\begin{pmatrix} -1/(b\cdot\tan\gamma)&-1/b&0\\1/(a\cdot\sin\gamma)&0&0\\0&0&1/c \end{pmatrix},\;
U=\begin{pmatrix}-1&0&0\\  1&0&0 \\0&-1&0\\ 0&1&0\\ 0&0&-1\\ \bluerow{0}{0}{1} \\ \redrow{0}{0}{1} \end{pmatrix},\; u=\begin{pmatrix} 0\\ 1/2\\ 0\\1\\ 0\\ \textcolor{blue}{1/2}\\ \textcolor{red}{1} \end{pmatrix}
\end{equation*}

\item $G=5$ [$\Gamma_{m}^{b}$]
\begin{equation*}
\begin{pmatrix}\bm{G}_{1}^{\text{ref}}\\\bm{G}_{2}^{\text{ref}}\\\bm{G}_{3}^{\text{ref}} \end{pmatrix}=\begin{pmatrix} -1/(b\cdot\tan\gamma)&-1/b&0\\1/(a\cdot\sin\gamma)&0&-1/c\\1/(a\cdot\sin\gamma)&0&1/c \end{pmatrix},\;\;
U,u \text{ as in }G=3
\end{equation*}
\end{itemize}

\item $F=C_{s}$ ($\sigma_{z}$)
\begin{equation*}
xx;xx;xx=2,\; yy;yy;yy=2,\; xx;yy;yy=2,\; yy;xx;xx=2,\; xx;zz;zz=2,\; yy;zz;zz=2
\end{equation*}
\begin{equation*}
\sigma^{x;xx},\; \sigma^{y;yy},\; \sigma^{x;yy},\; \sigma^{y;xx},\; \sigma^{x;xy},\; \sigma^{y;yx},\; \sigma^{x;zz},\; \sigma^{y;zz},\; \sigma^{z;zx},\; \sigma^{z;zy}
\end{equation*}
\begin{itemize}
\item $G=6-7$ [$\Gamma_{m}$]
\begin{equation*}
U=\begin{pmatrix}-1&0&0\\  \bluerow{1}{0}{0}\\ \redrow{1}{0}{0} \\0&-1&0\\ 0&1&0\\ 0&0&-1\\ 0&0&1  \end{pmatrix},\; u=\begin{pmatrix} 0\\ \textcolor{blue}{1/2}\\ \textcolor{red}{1}\\ 0\\1\\ 0\\ 1/2 \end{pmatrix}
\end{equation*}

\item $G=8-9$ [$\Gamma_{m}^{b}$]
\begin{equation*}
U,u \text{ as in }G=6
\end{equation*}
\end{itemize}
      
\item $F=D_{2}$
\begin{equation*}
xx;yy;zz=4
\end{equation*}
\begin{equation*}
\sigma^{x;yz},\; \sigma^{y;zx},\; \sigma^{z;xy}
\end{equation*}
\begin{itemize}
\item $G=16-19$ [$\Gamma_{o}$]
\begin{equation*}
\begin{pmatrix}\bm{G}_{1}^{\text{ref}}\\\bm{G}_{2}^{\text{ref}}\\\bm{G}_{3}^{\text{ref}} \end{pmatrix}=\begin{pmatrix} 0&-1/b&0\\1/a&0&0\\0&0&1/c \end{pmatrix},\;
U=\begin{pmatrix} 1&0&0\\-1&0&0\\ 0&-1&0\\0&1&0\\ \bluerow{0}{0}{-1}\\0&0&1 \\ \redrow{0}{0}{-1}\end{pmatrix},\; u=\begin{pmatrix} 0\\1/2\\ 0\\1/2\\\textcolor{blue}{0}\\1/2\\ \textcolor{red}{1/2} \end{pmatrix}
\end{equation*}

\item $G=20-21$ [$\Gamma_{o}^{b}$]. $\eta\equiv\frac{b^{2}-a^{2}}{b^{2}+a^{2}}$
\begin{equation*}
\begin{pmatrix}\bm{G}_{1}^{\text{ref}}\\\bm{G}_{2}^{\text{ref}}\\\bm{G}_{3}^{\text{ref}} \end{pmatrix}=\begin{pmatrix} 1/a&-1/b&0\\1/a&1/b&0\\0&0&1/c \end{pmatrix} \text{with }a>b,\;
U=\begin{pmatrix} 1&1&0\\ \eta&1&0\\ 0&0&1 \\ 1&-1&0\\ -1&-1&0\\ \bluerow{0}{0}{-1}\\ \redrow{0}{0}{-1} \end{pmatrix},\;
u=\begin{pmatrix} 1\\1/2\\1/2\\0\\ 0\\ \textcolor{blue}{0}\\  \textcolor{red}{1/2} \end{pmatrix}
\end{equation*}

\item $G=22$ [$\Gamma_{o}^{f}$]
\begin{equation*}
\begin{pmatrix}\bm{G}_{1}^{\text{ref}}\\\bm{G}_{2}^{\text{ref}}\\\bm{G}_{3}^{\text{ref}} \end{pmatrix}=\begin{pmatrix} 1/a&1/b&1/c\\-1/a&-1/b&1/c\\1/a&-1/b&-1/c \end{pmatrix}
\end{equation*}
\begin{itemize}
\item Variation 1: $a^{-2}\geq b^{-2}+c^{-2}$. $\mu_{ab}\equiv\frac{-a^{2}b^{2}+a^{2}c^{2}+b^{2}c^{2}}{a^{2}b^{2}+a^{2}c^{2}+b^{2}c^{2}}$, $\mu_{ac}\equiv\frac{a^{2}b^{2}-a^{2}c^{2}+b^{2}c^{2}}{a^{2}b^{2}+a^{2}c^{2}+b^{2}c^{2}}$
\begin{equation*}
U=\begin{pmatrix} 1&-\mu_{ab}&\mu_{ab}+\mu_{ac}-1\\ 1&-1&-1\\1&1&-1\\-1&1&1\\-1&1&-1\\ \bluerow{-1}{-1}{1}\\ \redrow{-1}{-1}{1} \\ \redrow{\mu_{ab}}{-1}{\mu_{ac}}   \end{pmatrix},\;
u=\begin{pmatrix} 1/2\\1\\1\\0\\0\\ \textcolor{blue}{0}\\ \textcolor{red}{1} \\ \textcolor{red}{1/2}  \end{pmatrix}
\end{equation*}
\item Variation 2: $a^{-2}<b^{-2}+c^{-2}$ and $b^{-2}<c^{-2}+a^{-2}$. $\mu_{ab}\equiv\frac{-a^{2}b^{2}+a^{2}c^{2}+b^{2}c^{2}}{a^{2}b^{2}+a^{2}c^{2}+b^{2}c^{2}}$, $\mu_{ac}\equiv\frac{a^{2}b^{2}-a^{2}c^{2}+b^{2}c^{2}}{a^{2}b^{2}+a^{2}c^{2}+b^{2}c^{2}}$
\begin{equation*}
U=\begin{pmatrix} 1&-\mu_{ab}&\mu_{ab}+\mu_{ac}-1\\ 1&-1&-1\\1&1&-1\\ 1&-1&1\\-1&1&1\\-1&1&-1\\ \bluerow{-1}{-1}{1}\\ \redrow{-1}{-1}{1}\\ \redrow{\mu_{ab}}{-1}{\mu_{ac}}   \end{pmatrix},\;
u=\begin{pmatrix} 1/2\\1\\1\\1\\0\\0\\ \textcolor{blue}{0}\\ \textcolor{red}{1}\\ \textcolor{red}{1/2}  \end{pmatrix}
\end{equation*}
\end{itemize}

\item $G=23-24$ [$\Gamma_{o}^{v}$]. $\eta_{ab}\equiv\frac{a^{2}}{a^{2}+b^{2}}$, $\eta_{ac}\equiv\frac{a^{2}}{a^{2}+c^{2}}$, $\eta_{bc}\equiv\frac{b^{2}}{b^{2}+c^{2}}$
\begin{equation*}
\begin{pmatrix}\bm{G}_{1}^{\text{ref}}\\\bm{G}_{2}^{\text{ref}}\\\bm{G}_{3}^{\text{ref}}\end{pmatrix}=\begin{pmatrix} 1/a&0&1/c\\0&-1/b&1/c\\1/a&-1/b&0\end{pmatrix} \text{with }c=\max(a,b,c)\:,
\end{equation*}\begin{equation*}
U=\begin{pmatrix} 1&\eta_{ac}&1-\eta_{ac}\\ \eta_{bc}&2\eta_{bc}-1&\eta_{bc}-1\\ 1-\eta_{ab}&-\eta_{ab}&1-2\eta_{ab}\\1&1&0\\0&1&1\\ -1&0&-1\\ \bluerow{-1}{-1}{0}\\ \redrow{-1}{-1}{0}\\ \redrow{1-2\eta_{ac}}{-\eta_{ac}}{1-\eta_{ac}}\\ \redrow{-\eta_{bc}}{-1}{\eta_{bc}-1} \end{pmatrix},\;
u=\begin{pmatrix} 1/2\\1/2\\1/2\\1\\0\\0\\ \textcolor{blue}{0}\\ \textcolor{red}{1}\\ \textcolor{red}{1/2}\\ \textcolor{red}{1/2} \end{pmatrix}
\end{equation*}
\end{itemize}

\item $F=C_{2v}$ ($c_{2,z}$, $\sigma_{x}$)
\begin{equation*}
zz;zz;zz=4,\; zz;xx;xx=4,\; zz;yy;yy=4
\end{equation*}
\begin{equation*}
\sigma^{z;zz},\; \sigma^{z;xx},\; \sigma^{z;yy},\; \sigma^{x;xz},\; \sigma^{y;yz}
\end{equation*}
\begin{itemize}
\item $G=25-34$ [$\Gamma_{o}$] $\rightarrow$ Same as $G=16$
\item $G=35-41$ [$\Gamma_{o}^{b}$] $\rightarrow$ Same as $G=20$
\item $G=42-43$ [$\Gamma_{o}^{f}$] $\rightarrow$ Same as $G=22$
\item $G=44-46$ [$\Gamma_{o}^{v}$] $\rightarrow$ Same as $G=23$
\end{itemize}

\item $F=C_{4}$ ($c_{4,z}$)
\begin{equation*}
zz;zz;zz=4,\; zz;xx;xx=zz;yy;yy=zz;xy;xy=2,\; xx;yy;zz=-xy;yx;zz=2
\end{equation*}
\begin{equation*}
\sigma^{z;zz},\; \sigma^{z;xx}=\sigma^{z;yy},\; \sigma^{x;xz}=\sigma^{y;yz},\; \sigma^{x;yz}=-\sigma^{y;zx},\; \Im\sigma^{z;xy}_{I}
\end{equation*}
\begin{itemize}
\item $G=75-78$ [$\Gamma_{q}$]
\begin{equation*}
\begin{pmatrix}\bm{G}_{1}^{\text{ref}}\\\bm{G}_{2}^{\text{ref}}\\\bm{G}_{3}^{\text{ref}} \end{pmatrix}=\begin{pmatrix} 1/a&0&0\\0&1/a&0\\0&0&1/c \end{pmatrix},\;
U=\begin{pmatrix} -1&0&0\\1&0&0\\0&-1&0\\0&1&0\\0&0&1\\ \bluerow{0}{0}{-1}\\ \redrow{0}{0}{-1}\end{pmatrix},\; 
u=\begin{pmatrix} 0\\1/2\\0\\1/2\\1/2\\ \textcolor{blue}{0} \\ \textcolor{red}{1/2}  \end{pmatrix}
\end{equation*}

\item $G=79-80$ [$\Gamma_{q}^{v}$]
\begin{equation*}
\begin{pmatrix}\bm{G}_{1}^{\text{ref}}\\\bm{G}_{2}^{\text{ref}}\\\bm{G}_{3}^{\text{ref}} \end{pmatrix}=\begin{pmatrix} 0&1/a&1/c\\1/a&0&1/c\\1/a&1/a&0 \end{pmatrix}\end{equation*}
\begin{itemize}
\item Variation 1: $a>c$. $\eta_{ac}\equiv\frac{a^{2}}{a^{2}+c^{2}}$
\begin{equation*}
U=\begin{pmatrix}\eta_{ac}&1&1-\eta_{ac}\\1&1&2\\-1&0&-1\\0&-1&-1\\1&\eta_{ac}&1-\eta_{ac}\\ \bluerow{-1}{-1}{0}\\ \redrow{-\eta_{ac}}{1-2\eta_{ac}}{1-\eta_{ac}}\\ \redrow{1-2\eta_{ac}}{-\eta_{ac}}{1-\eta_{ac}}   \end{pmatrix},\;
u=\begin{pmatrix}1/2\\1\\0\\0\\1/2\\ \textcolor{blue}{0}\\ \textcolor{red}{1/2}\\ \textcolor{red}{1/2}  \end{pmatrix}
\end{equation*}
\item Variation 2: $c>a$. $\eta_{ac}\equiv\frac{a^{2}}{a^{2}+c^{2}}$
\begin{equation*}
U=\begin{pmatrix}\eta_{ac}&1&1-\eta_{ac}\\1&1&2\\ 1&1&0\\-1&0&-1\\0&-1&-1\\1&\eta_{ac}&1-\eta_{ac}\\ \bluerow{-1}{-1}{0}\\ \redrow{-1}{-1}{0}\\ \redrow{-\eta_{ac}}{1-2\eta_{ac}}{1-\eta_{ac}}\\ \redrow{1-2\eta_{ac}}{-\eta_{ac}}{1-\eta_{ac}}   \end{pmatrix},\;
u=\begin{pmatrix}1/2\\1\\1\\0\\0\\1/2\\ \textcolor{blue}{0}\\ \textcolor{red}{1}\\ \textcolor{red}{1/2}\\ \textcolor{red}{1/2}  \end{pmatrix}
\end{equation*}
\end{itemize}
\end{itemize}

\item $F=S_{4}$ ($s_{4,z}$)
\begin{equation*}
zz;xx;xx=zz;yy;yy=-zz;xy;xy=2,\; xx;yy;zz=xy;yx;zz=2 
\end{equation*}
\begin{equation*}
\sigma^{z;xx}=-\sigma^{z;yy},\; \sigma^{x;xz}=-\sigma^{y;yz},\; \sigma^{x;yz}=\sigma^{y;zx},\; \sigma^{z;xy}_{I}=\sigma^{z;xy}_{II} 
\end{equation*}
\begin{itemize}
\item $G=81$ [$\Gamma_{q}$] $\rightarrow$ Same as $G=75$
\item $G=82$ [$\Gamma_{q}^{v}$] $\rightarrow$ Same as $G=79$
\end{itemize}

\item $F=D_{4}$ ($c_{4,z}$, $c_{2,x}$)
\begin{equation*}
xx;yy;zz=-xy;yx;zz=4,\; 
\end{equation*}
\begin{equation*}
\sigma^{x;yz}=-\sigma^{y;zx},\; \Im\sigma^{z;xy}_{I}
\end{equation*}
\begin{itemize}
\item $G=89-96$ [$\Gamma_{q}$]
\begin{equation*}
U=\begin{pmatrix} 1&-1&0\\-1&0&0\\ \bluerow{0}{0}{-1}\\0&0&1\\0&1&0 \\ \redrow{0}{0}{-1}\end{pmatrix},\;
u=\begin{pmatrix} 0\\0\\ \textcolor{blue}{0}\\1/2\\1/2\\ \textcolor{red}{1/2}  \end{pmatrix}
\end{equation*}
\item $G=97-98$ [$\Gamma_{q}^{v}$]
\begin{itemize}
\item Variation 1: $a> c$. $\eta_{ac}\equiv\frac{a^{2}}{a^{2}+c^{2}}$
\begin{equation*}
U=\begin{pmatrix}\eta_{ac}&1&1-\eta_{ac}\\1&1&2\\1&-1&0\\-1&0&-1\\ \bluerow{-1}{-1}{0}\\ \redrow{-\eta_{ac}}{1-2\eta_{ac}}{1-\eta_{ac}}   \end{pmatrix},\;
u=\begin{pmatrix}1/2\\1\\0\\0\\ \textcolor{blue}{0}\\ \textcolor{red}{1/2}  \end{pmatrix}
\end{equation*}
\item Variation 2: $c>a$. $\eta_{ac}\equiv\frac{a^{2}}{a^{2}+c^{2}}$
\begin{equation*}
U=\begin{pmatrix}\eta_{ac}&1&1-\eta_{ac}\\1&1&2\\1&1&0\\1&-1&0\\-1&0&-1\\ \bluerow{-1}{-1}{0}\\ \redrow{-\eta_{ac}}{1-2\eta_{ac}}{1-\eta_{ac}} \\ \redrow{-1}{-1}{0} \end{pmatrix},\;
u=\begin{pmatrix}1/2\\1\\1\\0\\0\\ \textcolor{blue}{0}\\ \textcolor{red}{1/2}\\ \textcolor{red}{1}  \end{pmatrix}
\end{equation*}
\end{itemize}
\end{itemize}

\item $F=C_{4v}$ ($c_{4,z}$, $\sigma_{x}$)
\begin{equation*}
zz;zz;zz=8,\; zz;xx;xx=zz;yy;yy=zz;xy;xy=4 
\end{equation*}
\begin{equation*}
\sigma^{z;zz},\; \sigma^{z;xx}=\sigma^{z;yy},\; \sigma^{x;xz}=\sigma^{y;yz} 
\end{equation*}
\begin{itemize}
\item $G=99-106$ [$\Gamma_{q}$] $\rightarrow$ Same as $G=89$
\item $G=107-110$ [$\Gamma_{q}^{v}$] $\rightarrow$ Same as $G=97$
\end{itemize}

\item $F=D_{2d}$ ($s_{4,z}$, $c_{2,x}$)
\begin{equation*}
xx;yy;zz=xy;yx;zz=4,\;
\end{equation*}
\begin{equation*}
\sigma^{x;yz}=\sigma^{y;zx},\; \sigma^{z;xy}_{I}=\sigma^{z;xy}_{II}
\end{equation*}
\begin{itemize}
\item $G=111-118$ [$\Gamma_{q}$] $\rightarrow$ Same as $G=89$
\item $G=119-122$ [$\Gamma_{q}^{v}$] $\rightarrow$ Same as $G=97$
\end{itemize}

\item $F=C_{3}$ ($c_{3,z}$)
\begin{equation*}\begin{aligned}
&xx;xx;xx=-xx;xy;xy=3/4,\; yy;yy;yy=-yy;yx;yx=3/4,\; zz;zz;zz=3,\; \\ &zz;xx;xx=zz;yy;yy=zz;xy;xy=3/2,\; xx;yy;zz=-xy;yx;zz=3/2
\end{aligned}\end{equation*}
\begin{equation*}\begin{aligned}
&\sigma^{x;xx}=-\sigma^{x;yy}=-\sigma^{y;yx}_{I}=-\sigma^{y;yx}_{II},\; \sigma^{y;yy}=-\sigma^{y;xx}=-\sigma^{x;xy}_{I}=-\sigma^{x;xy}_{II},\; \sigma^{z;zz},\; \sigma^{z;xx}=\sigma^{z;yy},\; \\ &\sigma^{x;xz}=\sigma^{y;yz},\; \sigma^{x;yz}=-\sigma^{y;zx},\; \Im\sigma^{z;xy}_{I}
\end{aligned}\end{equation*}
\begin{itemize}
\item $G=143-145$ [$\Gamma_{h}$]
\begin{equation*}
\begin{pmatrix}\bm{G}_{1}^{\text{ref}}\\\bm{G}_{2}^{\text{ref}}\\\bm{G}_{3}^{\text{ref}}\end{pmatrix}=\begin{pmatrix} 1/\sqrt{3}a&-1/a&0\\2/\sqrt{3}a&0&0\\0&0&1/c\end{pmatrix},\;
U=\begin{pmatrix}1&0&0\\0&0&-1\\0&0&1\\1&2&0\\ -1&1&0\\ \bluerow{-1}{-1}{0}\\ \redrow{0}{-1}{0}\\ \redrow{-2}{-1}{0}  \end{pmatrix},\;
u=\begin{pmatrix}0\\1/2\\1/2\\1\\ 1 \\ \textcolor{blue}{0}\\ \textcolor{red}{0}\\ \textcolor{red}{1} \end{pmatrix}
\end{equation*}
\item $G=146$ [$\Gamma_{rh}$]
\begin{equation*}
\begin{pmatrix}\bm{G}_{1}^{\text{ref}}\\\bm{G}_{2}^{\text{ref}}\\\bm{G}_{3}^{\text{ref}}\end{pmatrix}=\begin{pmatrix} 0&-2/3a&1/3c\\1/\sqrt{3}a&1/3a&1/3c\\-1/\sqrt{3}a&1/3a&1/3c\end{pmatrix}
\end{equation*}
\begin{itemize}
\item Variation 1: $a>\sqrt{2}c$. $\eta_{ac}\equiv\frac{a^{2}-2c^{2}}{a^{2}+c^{2}}$, $\eta_{a4c}\equiv\frac{2a^{2}-4c^{2}}{a^{2}+4c^{2}}$
\begin{equation*}
U=\begin{pmatrix} \eta_{a4c}&2&\eta_{a4c}\\0&1&-1\\-\eta_{a4c}&-\eta_{a4c}&-2\\ \bluerow{-1}{0}{1}\\1&-1&0\\ \redrow{-2}{-\eta_{a4c}}{-\eta_{a4c}}\\ \redrow{-1}{1}{0}\\ \redrow{0}{-1}{1}  \end{pmatrix},\;
u=\begin{pmatrix} 1\\1\\1\\ \textcolor{blue}{0}\\0\\ \textcolor{red}{1}\\ \textcolor{red}{1}\\ \textcolor{red}{0}     \end{pmatrix}
\end{equation*}
\item Variation 2: $a<\sqrt{2}c$. $\eta_{ac}\equiv\frac{a^{2}-2c^{2}}{a^{2}+c^{2}}$, $\eta_{a4c}\equiv\frac{2a^{2}-4c^{2}}{a^{2}+4c^{2}}$
\begin{equation*}
U=\begin{pmatrix} \eta_{a4c}&2&\eta_{a4c}\\1&1&\eta_{ac}\\-\eta_{a4c}&-\eta_{a4c}&-2\\ \bluerow{-1}{0}{1}\\ 1&1&1\\1&-1&0\\ \redrow{\eta_{ac}}{1}{1}\\ \redrow{-2}{-\eta_{a4c}}{-\eta_{a4c}}\\ \redrow{0}{-1}{1}\\-1&-1&-1\\-1&-\eta_{ac}&-1  \end{pmatrix},\;
u=\begin{pmatrix} 1\\1\\1\\ \textcolor{blue}{0}\\3/2\\0\\ \textcolor{red}{1}\\ \textcolor{red}{1}\\ \textcolor{red}{0}\\3/2\\1   \end{pmatrix}
\end{equation*}
\end{itemize}
\end{itemize}

\item $F=D_{3}$ ($c_{3,z}$, $c_{2,x}$)
\begin{equation*}
xx;xx;xx=-xx;xy;xy=3/2,\; xx;yy;zz=-xy;yx;zz=3
\end{equation*}
\begin{equation*}
\sigma^{x;xx}=-\sigma^{x;yy}=-\sigma^{y;yx}_{I}=-\sigma^{y;yx}_{II},\; \sigma^{x;yz}=-\sigma^{y;zx},\; \Im\sigma^{z;xy}_{I}
\end{equation*}
\begin{itemize}
\item $G=149-154$ [$\Gamma_{h}$]
\begin{equation*}
U=\begin{pmatrix}1&0&0\\0&0&-1\\0&0&1\\1&2&0\\ \bluerow{-2}{-1}{0}\\ \redrow{-1}{1}{0}\\ \redrow{-1}{-1}{0}  \end{pmatrix},\;
u=\begin{pmatrix}0\\1/2\\1/2\\1\\ \textcolor{blue}{0}\\ \textcolor{red}{1} \\ \textcolor{red}{0} \end{pmatrix}
\end{equation*}
\item $G=155$ [$\Gamma_{rh}$]
\begin{itemize}
\item Variation 1: $a>\sqrt{2}c$. $\eta_{ac}\equiv\frac{a^{2}-2c^{2}}{a^{2}+c^{2}}$, $\eta_{a4c}\equiv\frac{2a^{2}-4c^{2}}{a^{2}+4c^{2}}$
\begin{equation*}
U=\begin{pmatrix} \eta_{a4c}&2&\eta_{a4c}\\0&1&-1\\ \redrow{-\eta_{a4c}}{-\eta_{a4c}}{-2}\\ -1&0&1\\1&-1&0\\ \bluerow{-1}{-1}{-1} \end{pmatrix},\;
u=\begin{pmatrix} 1\\1\\ \textcolor{red}{1}\\0\\0\\ \textcolor{blue}{0} \end{pmatrix}
\end{equation*}
\item Variation 2: $a<\sqrt{2}c$. $\eta_{ac}\equiv\frac{a^{2}-2c^{2}}{a^{2}+c^{2}}$, $\eta_{a4c}\equiv\frac{2a^{2}-4c^{2}}{a^{2}+4c^{2}}$
\begin{equation*}
U=\begin{pmatrix} \eta_{a4c}&2&\eta_{a4c}\\1&1&\eta_{ac}\\-\eta_{a4c}&-\eta_{a4c}&-2\\ -1&0&1\\ 1&1&1\\1&-1&0\\ \redrow{-1}{-1}{-1}\\ \redrow{-1}{-\eta_{ac}}{-1}\\ \bluerow{-1}{-1}{-1}   \end{pmatrix},\;
u=\begin{pmatrix} 1\\1\\1\\ 0\\3/2\\0\\ \textcolor{red}{3/2}\\ \textcolor{red}{1}\\ \textcolor{blue}{0}   \end{pmatrix}
\end{equation*}
\end{itemize}
\end{itemize}

\item $F=C_{3v}$ ($c_{3,z}$, $\sigma_{y}$)
\begin{equation*}
xx;xx;xx=-xx;xy;xy=3/2,\; zz;zz;zz=6,\; zz;xx;xx=zz;yy;yy=zz;xy;xy=3
\end{equation*}
\begin{equation*}
\sigma^{x;xx}=-\sigma^{x;yy}=-\sigma^{y;yx}_{I}=-\sigma^{y;yx}_{II},\; \sigma^{z;zz},\; \sigma^{z;xx}=\sigma^{z;yy},\; \sigma^{x;xz}=\sigma^{y;yz}
\end{equation*}
\begin{itemize}
\item $G=156-159$ [$\Gamma_{h}$] $\rightarrow$ Same as $G=149$
\item $G=160-161$ [$\Gamma_{rh}$] $\rightarrow$ Same as $G=155$
\end{itemize}

\item $F=C_{6}$ ($c_{6,z}$)
\begin{equation*}zz;zz;zz=6,\; zz;xx;xx=zz;yy;yy=zz;xy;xy=3,\; xx;yy;zz=-xy;yx;zz=3 \end{equation*}
\begin{equation*}
\sigma^{z;zz},\; \sigma^{z;xx}=\sigma^{z;yy},\; \sigma^{x;xz}=\sigma^{y;yz},\; \sigma^{x;yz}=-\sigma^{y;zx},\; \Im\sigma^{z;xy} \end{equation*}
\begin{itemize}
\item $G=168-173$ [$\Gamma_{h}$]
\begin{equation*}
U=\begin{pmatrix}1&0&0\\0&0&1\\ \bluerow{0}{0}{-1}\\ \redrow{0}{0}{-1}\\1&2&0\\-1&1&0\\-1&-1&0  \end{pmatrix},\;
u=\begin{pmatrix}0\\1/2\\ \textcolor{blue}{0}\\ \textcolor{red}{1/2}\\1\\ 1 \\ 0 \end{pmatrix}
\end{equation*}
\end{itemize}

\item $F=C_{3h}$ ($c_{3,z}$, $\sigma_{z}$)
\begin{equation*}xx;xx;xx=-xx;xy;xy=3/2,\; yy;yy;yy=-yy;yx;yx=3/2
\end{equation*}
\begin{equation*}
\sigma^{x;xx}=-\sigma^{x;yy}=-\sigma^{y;yx}_{I}=-\sigma^{y;yx}_{II},\; \sigma^{y;yy}=-\sigma^{y;xx}=-\sigma^{x;xy}_{I}=-\sigma^{x;xy}_{II} \end{equation*}
\begin{itemize}
\item $G=174$ [$\Gamma_{h}$]
\begin{equation*}
U=\begin{pmatrix}1&0&0\\0&0&-1\\0&0&1\\1&2&0\\ -1&1&0\\ \bluerow{-1}{-1}{0}\\ \redrow{0}{-1}{0}\\ \redrow{-2}{-1}{0}  \end{pmatrix},\;
u=\begin{pmatrix}0\\0\\1/2\\1\\ 1 \\ \textcolor{blue}{0}\\ \textcolor{red}{0}\\ \textcolor{red}{1} \end{pmatrix}
\end{equation*}
\end{itemize}

\item $F=D_{6}$ ($c_{6,z}$, $c_{2,x}$)
\begin{equation*}xx;yy;zz=-xy;yx;zz=6
\end{equation*}
\begin{equation*}
\sigma^{x;yz}=-\sigma^{y;zx},\; \Im\sigma^{z;xy}_{I} \end{equation*}
\begin{itemize}
\item $G=177-182$ [$\Gamma_{h}$]
\begin{equation*}
U=\begin{pmatrix}1&0&0\\ \redrow{0}{0}{-1} \\ \bluerow{0}{0}{-1} \\0&0&1\\1&2&0\\ -2&-1&0 \end{pmatrix},\;
u=\begin{pmatrix}0\\ \textcolor{red}{1/2}\\ \textcolor{blue}{0}\\1/2\\1\\ 0\end{pmatrix}
\end{equation*}
\end{itemize}

\item $F=C_{6v}$ ($c_{6,z}$, $\sigma_{x}$)
\begin{equation*}zz;zz;zz=12,\; zz;xx;xx=zz;yy;yy=zz;xy;xy=6,
\end{equation*}
\begin{equation*}
\sigma^{z;zz},\; \sigma^{z;xx}=\sigma^{z;yy},\; \sigma^{x;xz}=\sigma^{y;yz} \end{equation*}
\begin{itemize}
\item $G=183-186$ [$\Gamma_{h}$] $\rightarrow$ Same as $G=177$
\end{itemize}

\item $F=D_{3h}$ ($s_{3,z}$, $c_{2,x}$)
\begin{equation*}xx;xx;xx=-xx;xy;xy=3,\; 
\end{equation*}
\begin{equation*}
\sigma^{x;xx}=-\sigma^{x;yy}=-\sigma^{y;yx}_{I}=-\sigma^{y;yx}_{II} \end{equation*}
\begin{itemize}
\item $G=187-190$ [$\Gamma_{h}$] 
\begin{equation*}
U=\begin{pmatrix}1&0&0\\0&0&-1\\0&0&1\\1&2&0\\ \bluerow{-2}{-1}{0}\\ \redrow{-1}{1}{0}\\ \redrow{-1}{-1}{0}  \end{pmatrix},\;
u=\begin{pmatrix}0\\0\\1/2\\1\\ \textcolor{blue}{0}\\ \textcolor{red}{1} \\ \textcolor{red}{0} \end{pmatrix}
\end{equation*}
\end{itemize}

\item $F=T$
\begin{equation*}
xx;yy;zz=xz;yx;zy=4
\end{equation*}\begin{equation*}
\sigma^{x;yz}=\sigma^{y;zx}=\sigma^{z;xy}
\end{equation*}
\begin{itemize}
\item $G=195$ \& $198$ [$\Gamma_{c}$]
\begin{equation*}
\begin{pmatrix}\bm{G}_{1}^{\text{ref}}\\\bm{G}_{2}^{\text{ref}}\\\bm{G}_{3}^{\text{ref}}\end{pmatrix}=\begin{pmatrix} 1/a&0&0\\0&1/a&0\\0&0&1/a\end{pmatrix},\; 
U=\begin{pmatrix} 0&1&0\\ -1&0&1\\ \bluerow{0}{0}{-1}\\0&-1&1\\1&0&0\\ \redrow{0}{-1}{-1}\\ \redrow{-1}{0}{-1} \end{pmatrix},\; u=\begin{pmatrix} 1/2\\0\\ \textcolor{blue}{0}\\0\\1/2\\ \textcolor{red}{0}\\ \textcolor{red}{0}\end{pmatrix}
\end{equation*}

\item $G=196$ [$\Gamma_{c}^{f}$]
\begin{equation*}
\begin{pmatrix}\bm{G}_{1}^{\text{ref}}\\\bm{G}_{2}^{\text{ref}}\\\bm{G}_{3}^{\text{ref}}\end{pmatrix}=\begin{pmatrix} -1/a&1/a&1/a\\1/a&-1/a&1/a\\1/a&1/a&-1/a
\end{pmatrix},\; 
U=\begin{pmatrix} 1&1&1\\ 1&-1&1\\ \bluerow{-1}{-1}{1}\\ 1&0&-1\\ -1&1&1\\0&1&-1\\ \redrow{-1}{-1}{3}\\ \redrow{0}{-1}{0}\\ \redrow{-1}{0}{0}
\end{pmatrix},\;
u=\begin{pmatrix} 3/2\\1\\ \textcolor{blue}{0}\\0\\1\\0\\ \textcolor{red}{3/2}\\ \textcolor{red}{0}\\ \textcolor{red}{0}
\end{pmatrix}
\end{equation*}

\item $G=197$ \& $199$ [$\Gamma_{c}^{v}$]
\begin{equation*}
\begin{pmatrix}\bm{G}_{1}^{\text{ref}}\\\bm{G}_{2}^{\text{ref}}\\\bm{G}_{3}^{\text{ref}}\end{pmatrix}=\begin{pmatrix} 0&1/a&1/a\\1/a&0&1/a\\1/a&1/a&0\end{pmatrix},\; 
U=\begin{pmatrix} 1&1&2\\ 1&0&-1\\ \bluerow{-1}{-1}{0}\\ 0&1&-1\\ \redrow{-1}{-2}{-1}\\ \redrow{-2}{-1}{-1} \end{pmatrix},\; 
u=\begin{pmatrix} 1\\0\\ \textcolor{blue}{0}\\0\\ \textcolor{red}{0}\\ \textcolor{red}{0}\end{pmatrix}
\end{equation*}
\end{itemize}

\item $F=O$
\begin{equation*}
xx;yy;zz=xz;yx;zy=-xy;yx;zz=-xx;yz;zy=-xz;yy;zx=4
\end{equation*}
\begin{equation*} \Im\sigma^{x;yz}_{I}=\Im\sigma^{y;zx}_{I}=\Im\sigma^{z;xy}_{I} \end{equation*}
\begin{itemize}
\item $G=207-208$ \& $212-213$ [$\Gamma_{c}$]
\begin{equation*}
U=\begin{pmatrix}1&-1&0\\ 0&1&0\\ -1&0&1\\ \bluerow{0}{0}{-1}\\ \redrow{-1}{0}{-1} \end{pmatrix},\; u=\begin{pmatrix}0\\ 1/2\\0\\ \textcolor{blue}{0}\\ \textcolor{red}{0}\end{pmatrix}
\end{equation*}

\item $G=209-210$ [$\Gamma_{c}^{f}$]
\begin{equation*}
U=\begin{pmatrix}-1&1&0\\ 1&1&1\\ 1&-1&1\\ \bluerow{-1}{-1}{1}\\ 1&0&-1\\ \redrow{-1}{-1}{3}\\ \redrow{0}{-1}{0}
\end{pmatrix},\; 
u=\begin{pmatrix}0\\ 3/2\\1\\ \textcolor{blue}{0}\\0\\ \textcolor{red}{3/2}\\ \textcolor{red}{0}
\end{pmatrix}
\end{equation*}

\item $G=211$ \& $214$ [$\Gamma_{c}^{v}$]
\begin{equation*}
U=\begin{pmatrix}-1&1&0\\ 1&1&2\\ 1&0&-1\\ \bluerow{-1}{-1}{0}\\ \redrow{-1}{-2}{-1} \end{pmatrix},\; 
u=\begin{pmatrix}0\\ 1\\0\\ \textcolor{blue}{0}\\ \textcolor{red}{0}\end{pmatrix}
\end{equation*}
\end{itemize}

\item $F=T_{d}$
\begin{equation*}
xx;yy;zz=xz;yx;zy=xy;yx;zz=xx;yz;zy=xz;yy;zx=4
\end{equation*}
\begin{equation*}
\sigma^{x;yz}_{I}=\sigma^{x;yz}_{II}=\sigma^{y;zx}_{I}=\sigma^{y;zx}_{II}=\sigma^{z;xy}_{I}=\sigma^{z;xy}_{II}
\end{equation*}

\begin{itemize}
\item $G=215$ \& $218$ [$\Gamma_{c}$] $\rightarrow$ Same as $G=207$
\item $G=216$ \& $219$ [$\Gamma_{c}^{f}$] $\rightarrow$ Same as $G=209$
\item $G=217$ \& $220$ [$\Gamma_{c}^{v}$] $\rightarrow$ Same as $G=211$
\end{itemize}
      \vspace{0.1cm}
      \hrule\vspace{0.05cm}\hrule
\end{itemize}

The folded equation for magnetic point groups of type III, \eqref{sumIII}, can be readily computed from the previous list. Since the evaluation is slightly more involved, we next provide the explicit expressions for all non-vanishing components $\sigma^{a;bc}_{\text{shift}}$ in the non-trivial cases, namely for those groups $M$ where $H$ is non-centrosymmetric ($i\in H\Rightarrow\sigma^{a;bc}_{\text{shift}}=0$) or there is $\pazocal{P}\pazocal{T}$ symmetry ($i\in F-H\Rightarrow\mathbb{D}^{F}=0$), in which case \eqref{sumIII} reduces to  
\begin{equation*}
\sigma^{a;bc}_{\text{shift}}=-\frac{i}{\abs{H}N^{\text{IBZ}}_{\bm{k}}}\sum_{\bm{k}\in\text{IBZ}}\sum_{a',b',c'}\mathbb{D}^{H}_{a,a';b,b';c,c'}\Re\left(\tilde{\sigma}^{a';b'c'}_{\text{shift}}(\bm{k})-\tilde{\sigma}^{a';c'b'}_{\text{shift}}(\bm{k})\right)
\end{equation*}
which can be evaluated straightforwardly as \eqref{sumI} or \eqref{sumII}. The general form of the shift conductivity tensor can then be immediately identified from these expressions for any magnetic point group. The type III groups are labelled with an arbitrary number and in the Shubnikov-Belov notation as in Reference \cite{bradley2010mathematical}. The orientation of each ordinary point group $H$ and $F$ is as in the previous list, except when specified by new generators. For simplicity, we omit the ``shift'' label, the $\bm{k}$ dependence and the ubiquitous $\frac{1}{N^{\text{IBZ}}_{\bm{k}}}\sum_{\bm{k}\in\text{IBZ}}$ in the notation of the conductivity. The general relations $\sigma^{a;bc}_{\text{shift}}=(\sigma^{a;cb}_{\text{shift}})^{*}$ are also omitted.

\begin{itemize}
      \hrule\vspace{0.05cm}\hrule
      \vspace{-0.2cm}
\item $M=2$ $(2')$, $H=C_{1}$, $F=C_{2}$
\begin{equation*}
\sigma^{a;bc}=\left\{\begin{aligned}
&\Im\left[\tilde{\sigma}^{a;bc}+\tilde{\sigma}^{a;cb}\right], &\text{ if }{\#x+\#y} \text{ is even} \\ 
&-i\Re\left[\tilde{\sigma}^{a;bc}-\tilde{\sigma}^{a;cb}\right], &\text{ if }{\#x+\#y} \text{ is odd}
\end{aligned}\right.
\end{equation*}
where $\#x$ ($\#y$) is the number of times that $x$ ($y$, resp.) appears in the ($a,b,c$) triplet.

\item $M=3$ $(m')$, $H=C_{1}$, $F=C_{s}$
\begin{equation*}
\sigma^{a;bc}=\left\{\begin{aligned}
&\Im\left[\tilde{\sigma}^{a;bc}+\tilde{\sigma}^{a;cb}\right], &\text{ if }{\#x+\#y} \text{ is odd} \\ 
&-i\Re\left[\tilde{\sigma}^{a;bc}-\tilde{\sigma}^{a;cb}\right], &\text{ if }{\#x+\#y} \text{ is even}
\end{aligned}\right.
\end{equation*}
where $\#x$ ($\#y$) is the number of times that $x$ ($y$, resp.) appears in the ($a,b,c$) triplet.

\item $M=7$ $(2'2'2)$, $H=C_{2}$, $F=D_{2}$
\begin{equation*}\begin{aligned}
&\sigma^{x;xz}=-i\Re\left[\tilde{\sigma}^{x;xz}-\tilde{\sigma}^{x;zx}\right],\;
\sigma^{y;yz}=-i\Re\left[\tilde{\sigma}^{y;yz}-\tilde{\sigma}^{y;zy}\right],\; \\
&\sigma^{x;yz}=\Im\left[\tilde{\sigma}^{x;yz}+\tilde{\sigma}^{x;zy}\right],\; 
\sigma^{y;zx}=\Im\left[\tilde{\sigma}^{y;zx}+\tilde{\sigma}^{y;xz}\right],\;
\sigma^{z;xy}=\Im\left[\tilde{\sigma}^{z;xy}+\tilde{\sigma}^{z;yx}\right]
\end{aligned}\end{equation*}

\item $M=8$ $(m'm'2)$, $H=C_{2}$, $F=C_{2v}$
\begin{equation*}\begin{aligned}
&\sigma^{z;zz}=2\Im\tilde{\sigma}^{z;zz},\;
\sigma^{z;xx}=2\Im\tilde{\sigma}^{z;xx},\;
\sigma^{z;yy}=2\Im\tilde{\sigma}^{z;yy},\;  \\
&\sigma^{x;xz}=\Im\left[\tilde{\sigma}^{x;xz}+\tilde{\sigma}^{x;zx}\right],\;
\sigma^{y;yz}=\Im\left[\tilde{\sigma}^{y;yz}+\tilde{\sigma}^{y;zy}\right],\;
\sigma^{x;yz}=-i\Re\left[\tilde{\sigma}^{x;yz}-\tilde{\sigma}^{x;zy}\right],\; \\
&\sigma^{y;zx}=-i\Re\left[\tilde{\sigma}^{y;zx}-\tilde{\sigma}^{y;xz}\right],\;
\sigma^{z;xy}=-i\Re\left[\tilde{\sigma}^{z;xy}-\tilde{\sigma}^{z;yx}\right]
\end{aligned}\end{equation*}

\item $M=9$ $(m'm2')$, $H=C_{s}$ ($\sigma_{y}$), $F=C_{2v}$
\begin{equation*}\begin{aligned}
&\sigma^{z;zz}=2\Im\tilde{\sigma}^{z;zz},\;
\sigma^{z;xx}=2\Im\tilde{\sigma}^{z;xx},\;
\sigma^{z;yy}=2\Im\tilde{\sigma}^{z;yy},\;
\sigma^{y;yx}=-i\Re\left[\tilde{\sigma}^{y;yx}-\tilde{\sigma}^{y;xy}\right],\;  \\
&\sigma^{x;xz}=\Im\left[\tilde{\sigma}^{x;xz}+\tilde{\sigma}^{x;zx}\right],\;
\sigma^{y;yz}=\Im\left[\tilde{\sigma}^{y;yz}+\tilde{\sigma}^{y;zy}\right],\;
\sigma^{z;zx}=-i\Re\left[\tilde{\sigma}^{z;zx}-\tilde{\sigma}^{z;xz}\right]
\end{aligned}\end{equation*}

\item $M=13$ $(4')$, $H=C_{2}$, $F=C_{4}$
\begin{equation*}\begin{aligned}
&\sigma^{z;zz}=2\Im\tilde{\sigma}^{z;zz},\;
\sigma^{z;xx}=\sigma^{z;yy}=\Im\left[\tilde{\sigma}^{z;xx}+\tilde{\sigma}^{z;yy}\right],\; \\
&\sigma^{x;xz}=(\sigma^{y;yz})^{*}=-\frac{i}{2}\left[\tilde{\sigma}^{x;xz}-(\tilde{\sigma}^{x;zx})^{*}-(\tilde{\sigma}^{y;yz})^{*}+\tilde{\sigma}^{y;zy} \right],\; \\
&\sigma^{x;yz}=-\sigma^{y;zx}=-\frac{i}{2}\left[\tilde{\sigma}^{x;yz}-(\tilde{\sigma}^{x;zy})^{*}+(\tilde{\sigma}^{y;xz})^{*}-\tilde{\sigma}^{y;zx} \right]
\end{aligned}\end{equation*}

\item $M=14$ $(\overline{4}')$, $H=C_{2}$, $F=S_{4}$
\begin{equation*}\begin{aligned}
&\sigma^{z;xx}=-\sigma^{z;yy}=\Im\left[\tilde{\sigma}^{z;xx}-\tilde{\sigma}^{z;yy}\right],\; 
\sigma^{x;xz}=-(\sigma^{y;yz})^{*}=-\frac{i}{2}\left[\tilde{\sigma}^{x;xz}-(\tilde{\sigma}^{x;zx})^{*}+(\tilde{\sigma}^{y;yz})^{*}-\tilde{\sigma}^{y;zy} \right],\; \\
&\sigma^{x;yz}=\sigma^{y;zx}=-\frac{i}{2}\left[\tilde{\sigma}^{x;yz}-(\tilde{\sigma}^{x;zy})^{*}-(\tilde{\sigma}^{y;xz})^{*}+\tilde{\sigma}^{y;zx} \right],\;
\sigma^{z;xy}=-i\left[ \tilde{\sigma}^{z;xy}-(\tilde{\sigma}^{z;yx})^{*} \right]
\end{aligned}\end{equation*}

\item $M=15$ $(42'2')$, $H=C_{4}$, $F=D_{4}$
\begin{equation*}\begin{aligned}
&\sigma^{x;xz}=\sigma^{y;yz}=-\frac{i}{2}\Re\left[\tilde{\sigma}^{x;xz}-\tilde{\sigma}^{x;zx}+\tilde{\sigma}^{y;yz}-\tilde{\sigma}^{y;zy} \right],\; \\
&\sigma^{x;yz}=-\sigma^{y;zx}=\frac{1}{2}\Im\left[\tilde{\sigma}^{x;yz}+\tilde{\sigma}^{x;zy}-\tilde{\sigma}^{y;xz}-\tilde{\sigma}^{y;zx} \right]
\end{aligned}\end{equation*}

\item $M=16$ $(4'22')$, $H=D_{2}$, $F=D_{4}$
\begin{equation*}
\sigma^{x;yz}=-\sigma^{y;zx}=-\frac{i}{2}\left[\tilde{\sigma}^{x;yz}-(\tilde{\sigma}^{x;zy})^{*}+(\tilde{\sigma}^{y;xz})^{*}-\tilde{\sigma}^{y;zx} \right]
\end{equation*}

\item $M=20$ $(4m'm')$, $H=C_{4}$, $F=C_{4v}$
\begin{equation*}\begin{aligned}
&\sigma^{z;zz}=2\Im\tilde{\sigma}^{z;zz},\; \sigma^{z;xx}=\sigma^{z;yy}=\Im\left[\tilde{\sigma}^{z;xx}+\tilde{\sigma}^{z;yy} \right],\; \\  
&\sigma^{x;xz}=\sigma^{y;yz}=\frac{1}{2}\Im\left[\tilde{\sigma}^{x;xz}+\tilde{\sigma}^{x;zx}+\tilde{\sigma}^{y;yz}+\tilde{\sigma}^{y;zy} \right],\; \\ 
&\sigma^{x;yz}=\sigma^{y;zx}=-\frac{i}{2}\Re\left[\tilde{\sigma}^{x;yz}-\tilde{\sigma}^{x;zy}-\tilde{\sigma}^{y;xz}+\tilde{\sigma}^{y;zx} \right],\; 
\sigma^{z;xy}=-i\Re\left[ \tilde{\sigma}^{z;xy}-\tilde{\sigma}^{z;yx} \right]
\end{aligned}\end{equation*}

\item $M=21$ $(4'mm')$, $H=C_{2v}$, $F=C_{4v}$
\begin{equation*}\begin{aligned}
&\sigma^{z;zz}=2\Im\tilde{\sigma}^{z;zz},\; \sigma^{z;xx}=\sigma^{z;yy}=\Im\left[\tilde{\sigma}^{z;xx}+\tilde{\sigma}^{z;yy} \right],\; \\  
&\sigma^{x;xz}=(\sigma^{y;yz})^{*}=-\frac{i}{2}\left[\tilde{\sigma}^{x;xz}-(\tilde{\sigma}^{x;zx})^{*}-(\tilde{\sigma}^{y;yz})^{*}+\tilde{\sigma}^{y;zy} \right]
\end{aligned}\end{equation*}

\item $M=22$ $(\overline{4}2'm')$, $H=S_{4}$, $F=D_{2d}$
\begin{equation*}\begin{aligned}
&\sigma^{x;xz}=-\sigma^{y;yz}=-\frac{i}{2}\Re\left[\tilde{\sigma}^{x;xz}-\tilde{\sigma}^{x;zx}-\tilde{\sigma}^{y;yz}+\tilde{\sigma}^{y;zy} \right],\; \\
&\sigma^{x;yz}=\sigma^{y;zx}=\frac{1}{2}\Im\left[\tilde{\sigma}^{x;yz}+\tilde{\sigma}^{x;zy}+\tilde{\sigma}^{y;xz}+\tilde{\sigma}^{y;zx}\right],\; 
\sigma^{z;xy}=\Im\left[\tilde{\sigma}^{z;xy}+\tilde{\sigma}^{z;yx}\right]
\end{aligned}\end{equation*}

\item $M=23$ $(\overline{4}'2m')$, $H=D_{2}$, $F=D_{2d}$
\begin{equation*}
\sigma^{x;yz}=\sigma^{y;zx}=-\frac{i}{2}\left[\tilde{\sigma}^{x;yz}-(\tilde{\sigma}^{x;zy})^{*}-(\tilde{\sigma}^{y;xz})^{*}+\tilde{\sigma}^{y;zx}\right],\; 
\sigma^{z;xy}=-i\left[\tilde{\sigma}^{z;xy}-(\tilde{\sigma}^{z;yx})^{*}\right]
\end{equation*}

\item $M=24$ $(\overline{4}'m2')$, $H=C_{2v}$, $F=D_{2d}$ ($s_{4,z}$, $\sigma_{x}$)
\begin{equation*}
\sigma^{z;xx}=-\sigma^{z;yy}=\Im\left[\tilde{\sigma}^{z;xx}-\tilde{\sigma}^{z;yy}\right],\; 
\sigma^{x;xz}=-(\sigma^{y;yz})^{*}=-\frac{i}{2}\left[\tilde{\sigma}^{x;xz}-(\tilde{\sigma}^{x;zx})^{*}+(\tilde{\sigma}^{y;yz})^{*}-\tilde{\sigma}^{y;zy}\right]
\end{equation*}

\item $M=30$ $(32')$, $H=C_{3}$, $F=D_{3}$
\begin{equation*}\begin{aligned}
&\sigma^{x;xx}=-\sigma^{x;yy}=-\sigma^{y;yx}=\frac{1}{2}\Im\left[\tilde{\sigma}^{x;xx}-\tilde{\sigma}^{x;yy}-\tilde{\sigma}^{y;yx}-\tilde{\sigma}^{y;xy}\right],\; \\
&\sigma^{x;xz}=\sigma^{y;yz}=-\frac{i}{2}\Re\left[ \tilde{\sigma}^{x;xz}-\tilde{\sigma}^{x;zx}+\tilde{\sigma}^{y;yz}-\tilde{\sigma}^{y;zy} \right],\; \\ 
&\sigma^{x;yz}=-\sigma^{y;zx}=\frac{1}{2}\Im\left[\tilde{\sigma}^{x;yz}+\tilde{\sigma}^{x;zy}-\tilde{\sigma}^{y;xz}-\tilde{\sigma}^{y;zx} \right]
\end{aligned}\end{equation*}

\item $M=31$ $(3m')$, $H=C_{3}$, $F=C_{3v}$
\begin{equation*}\begin{aligned}
&\sigma^{x;xx}=-\sigma^{x;yy}=-\sigma^{y;yx}=\frac{1}{2}\Im\left[\tilde{\sigma}^{x;xx}-\tilde{\sigma}^{x;yy}-\tilde{\sigma}^{y;yx}-\tilde{\sigma}^{y;xy}\right],\; 
\sigma^{z;zz}=2\Im\tilde{\sigma}^{z;zz},\; \\
&\sigma^{z;xx}=\sigma^{z;yy}=\Im\left[\tilde{\sigma}^{z;xx}+\tilde{\sigma}^{z;yy}\right],\; 
\sigma^{x;xz}=\sigma^{y;yz}=\frac{1}{2}\Im\left[ \tilde{\sigma}^{x;xz}+\tilde{\sigma}^{x;zx}+\tilde{\sigma}^{y;yz}+\tilde{\sigma}^{y;zy} \right],\; \\ 
&\sigma^{x;yz}=\sigma^{y;zx}=-\frac{i}{2}\Re\left[\tilde{\sigma}^{x;yz}-\tilde{\sigma}^{x;zy}-\tilde{\sigma}^{y;xz}+\tilde{\sigma}^{y;zx} \right],\; 
\sigma^{z;xy}=-i\Re\left[\tilde{\sigma}^{z;xy}-\tilde{\sigma}^{z;yx} \right]
\end{aligned}\end{equation*}

\item $M=32$ $(\overline{6}')$, $H=C_{3}$, $F=C_{3h}$
\begin{equation*}\begin{aligned}
&\sigma^{x;xx}=-\sigma^{x;yy}=-\sigma^{y;yx}=\frac{1}{2}\Im\left[\tilde{\sigma}^{x;xx}-\tilde{\sigma}^{x;yy}-\tilde{\sigma}^{y;yx}-\tilde{\sigma}^{y;xy}\right],\; \\
&\sigma^{y;yy}=-\sigma^{y;xx}=-\sigma^{x;xy}=\frac{1}{2}\Im\left[\tilde{\sigma}^{y;yy}-\tilde{\sigma}^{y;xx}-\tilde{\sigma}^{x;xy}-\tilde{\sigma}^{x;yx}\right],\; \\
&\sigma^{x;xz}=\sigma^{y;yz}=-\frac{i}{2}\Re\left[ \tilde{\sigma}^{x;xz}-\tilde{\sigma}^{x;zx}+\tilde{\sigma}^{y;yz}-\tilde{\sigma}^{y;zy} \right],\; \\ 
&\sigma^{x;yz}=\sigma^{y;zx}=-\frac{i}{2}\Re\left[\tilde{\sigma}^{x;yz}-\tilde{\sigma}^{x;zy}-\tilde{\sigma}^{y;xz}+\tilde{\sigma}^{y;zx} \right],\; 
\sigma^{z;xy}=-i\Re\left[\tilde{\sigma}^{z;xy}-\tilde{\sigma}^{z;yx} \right]
\end{aligned}\end{equation*}

\item $M=33$ $(\overline{6}m'2')$, $H=C_{3h}$, $F=D_{3h}$
\begin{equation*}
\sigma^{x;xx}=-\sigma^{x;yy}=-\sigma^{y;yx}=\frac{1}{2}\Im\left[\tilde{\sigma}^{x;xx}-\tilde{\sigma}^{x;yy}-\tilde{\sigma}^{y;yx}-\tilde{\sigma}^{y;xy}\right]
\end{equation*}

\item $M=34$ $(\overline{6}'m2')$, $H=C_{3v}$, $F=D_{3h}$
\begin{equation*}\begin{aligned}
&\sigma^{x;xx}=-\sigma^{x;yy}=-\sigma^{y;yx}=\frac{1}{2}\Im\left[\tilde{\sigma}^{x;xx}-\tilde{\sigma}^{x;yy}-\tilde{\sigma}^{y;yx}-\tilde{\sigma}^{y;xy}\right],\; \\
&\sigma^{x;xz}=\sigma^{y;yz}=-\frac{i}{2}\Re\left[\tilde{\sigma}^{x;xz}-\tilde{\sigma}^{x;zx}+\tilde{\sigma}^{y;yz}-\tilde{\sigma}^{y;zy} \right]
\end{aligned}\end{equation*}

\item $M=35$ $(\overline{6}'m'2)$, $H=D_{3}$, $F=D_{3h}$
\begin{equation*}\begin{aligned}
&\sigma^{x;xx}=-\sigma^{x;yy}=-\sigma^{y;yx}=\frac{1}{2}\Im\left[\tilde{\sigma}^{x;xx}-\tilde{\sigma}^{x;yy}-\tilde{\sigma}^{y;yx}-\tilde{\sigma}^{y;xy}\right],\; \\
&\sigma^{x;yz}=\sigma^{y;zx}=-\frac{i}{2}\Re\left[\tilde{\sigma}^{x;yz}-\tilde{\sigma}^{x;zy}-\tilde{\sigma}^{y;xz}+\tilde{\sigma}^{y;zx} \right],\;
\sigma^{z;xy}=-i\Re\left[\tilde{\sigma}^{z;xy}-\tilde{\sigma}^{z;yx}\right]
\end{aligned}\end{equation*}

\item $M=36$ $(6')$, $H=C_{3}$, $F=C_{6}$
\begin{equation*}\begin{aligned}
&\sigma^{z;zz}=2\Im\tilde{\sigma}^{z;zz},\;
\sigma^{z;xx}=\sigma^{z;yy}=\Im\left[\tilde{\sigma}^{z;xx}+\tilde{\sigma}^{z;yy}\right],\; \\
&\sigma^{x;xz}=\sigma^{y;yz}=\frac{1}{2}\Im\left[\tilde{\sigma}^{x;xz}+\tilde{\sigma}^{x;zx}+\tilde{\sigma}^{y;yz}+\tilde{\sigma}^{y;zy}\right],\; \\
&\sigma^{x;yz}=-\sigma^{y;zx}=\frac{1}{2}\Im\left[\tilde{\sigma}^{x;yz}+\tilde{\sigma}^{x;zy}-\tilde{\sigma}^{y;xz}-\tilde{\sigma}^{y;zx} \right]
\end{aligned}\end{equation*}

\item $M=41$ $(62'2')$, $H=C_{6}$, $F=D_{6}$
\begin{equation*}\begin{aligned}
&\sigma^{x;xz}=\sigma^{y;yz}=-\frac{i}{2}\Re\left[\tilde{\sigma}^{x;xz}-\tilde{\sigma}^{x;zx}+\tilde{\sigma}^{y;yz}-\tilde{\sigma}^{y;zy}\right],\; \\
&\sigma^{x;yz}=-\sigma^{y;zx}=\frac{1}{2}\Im\left[\tilde{\sigma}^{x;yz}+\tilde{\sigma}^{x;zy}-\tilde{\sigma}^{y;xz}-\tilde{\sigma}^{y;zx}\right]
\end{aligned}\end{equation*}

\item $M=42$ $(6'2'2)$, $H=D_{3}$, $F=D_{6}$
\begin{equation*}
\sigma^{x;yz}=-\sigma^{y;zx}=\frac{1}{2}\Im\left[\tilde{\sigma}^{x;yz}+\tilde{\sigma}^{x;zy}-\tilde{\sigma}^{y;xz}-\tilde{\sigma}^{y;zx}\right]
\end{equation*}

\item $M=46$ $(6m'm')$, $H=C_{6}$, $F=C_{6v}$
\begin{equation*}\begin{aligned}
&\sigma^{z;zz}=2\Im\tilde{\sigma}^{z;zz},\;
\sigma^{z;xx}=\sigma^{z;yy}=\Im\left[\tilde{\sigma}^{z;xx}+\tilde{\sigma}^{z;yy}\right],\; \\
&\sigma^{x;xz}=\sigma^{y;yz}=\frac{1}{2}\Im\left[\tilde{\sigma}^{x;xz}+\tilde{\sigma}^{x;zx}+\tilde{\sigma}^{y;yz}+\tilde{\sigma}^{y;zy}\right],\; \\
&\sigma^{x;yz}=\sigma^{y;zx}=-\frac{i}{2}\Re\left[\tilde{\sigma}^{x;yz}-\tilde{\sigma}^{x;zy}-\tilde{\sigma}^{y;xz}+\tilde{\sigma}^{y;zx} \right],\;
\sigma^{z;xy}=-i\Re\left[\tilde{\sigma}^{z;xy}-\tilde{\sigma}^{z;yx}\right]
\end{aligned}\end{equation*}

\item $M=47$ $(6'm'm)$, $H=C_{3v}$, $F=C_{6v}$
\begin{equation*}\begin{aligned}
&\sigma^{z;zz}=2\Im\tilde{\sigma}^{z;zz},\;
\sigma^{z;xx}=\sigma^{z;yy}=\Im\left[\tilde{\sigma}^{z;xx}+\tilde{\sigma}^{z;yy}\right],\; \\
&\sigma^{x;xz}=\sigma^{y;yz}=\frac{1}{2}\Im\left[\tilde{\sigma}^{x;xz}+\tilde{\sigma}^{x;zx}+\tilde{\sigma}^{y;yz}+\tilde{\sigma}^{y;zy}\right]
\end{aligned}\end{equation*}

\item $M=54$ $(\overline{4}'3m')$, $H=T$, $F=T_{d}$
\begin{equation*}
\sigma^{x;yz}=\sigma^{y;zx}=\sigma^{z;xy}=-\frac{i}{3}\left[\tilde{\sigma}^{x;yz}-(\tilde{\sigma}^{x;zy})^{*}+\tilde{\sigma}^{y;zx}-(\tilde{\sigma}^{y;xz})^{*}+\tilde{\sigma}^{z;xy}-(\tilde{\sigma}^{z;yx})^{*}\right]
\end{equation*}

\item $M=55$ $(4'32')$, $H=T$, $F=O$
\begin{equation*}
\sigma^{a;bc}=0,\;\forall a,b,c
\end{equation*}

      \vspace{0.1cm}
      \hrule\vspace{0.05cm}\hrule
\end{itemize}


\section{Conclusive remarks}
The use of Gaussian basis sets to compute the shift conductivity has been proved satisfactory. The analytical evaluation of real-space integrals involving localized functions allows to readily compute the Berry connection and velocity matrix elements, while the reduced dimension yields lighter calculations, and the economical option of hybrid functionals allows to easily reproduce the desired band gap with good precision in a wide variety of systems. Furthermore, the (magnetic) space group symmetry is fully preserved and it can be capitalised on to perform maximally-efficient reciprocal space summations, in addition to immediately discerning the contributions to the electrical current under any light polarization. 

The numerical results for the chosen materials are in standard agreement with the literature, and in all cases the length and velocity gauges have been shown to yield nearly identical outcomes. One would be tempted to conclude that the velocity gauge, in view of its comparative simplicity, should then be the preferred option in general. However, the length gauge makes no use of completeness relations and larger discrepancies may appear when employing smaller bases. In addition, care should be taken when separating the shift and injection contributions without time-reversal symmetry in the velocity gauge. Nevertheless, the use of as-largest-as-possible basis sets is generally advisable, typically between TZVP and QZVP and including diffuse exponents. The use of ghost atoms, i.e., basis functions not located on atomic positions, could be explored in order to facilitate the reproduction of particularly-delocalized empty conduction states.

We note that all other single-particle contributions to the BPVE (in particular, the injection current) and to the total second-order optical response (in particular, the second-harmonic generation) can be computed from this method since the corresponding expressions involve the same basic quantities as the shift current. The transformation properties of these other optical contributions are the same for spatial operations, but the role of time-reversal may change. For example, the injection conductivity has the opposite behaviour to $\sigma_{\text{shift}}$, in the sense that time-reversal symmetry forces it to be imaginary (instead of real); and the results in Section \ref{SecIBZ} can be adapted from that. The evaluation of metallic systems is also possible in the length gauge, albeit it introduces additional terms with Fermi surface derivatives \cite{Pedersen2017}.

\section*{Supporting Information}
Supporting Information available:  

Input files for the self-consistent electronic structure calculations in \texttt{CRYSTAL23} for each material (.d12 files, in the terminology of the code). Band structures along high-symmetry lines for each material. Explicit comparison between the BZ and IBZ summations in BaTiO$_{3}$.


\begin{acknowledgement}
The authors acknowledge financial support from Spanish MICINN (Grant Nos. PID2019-109539GB-C43 \& TED2021-131323B-I00 \& PID2022-141712NB-C21), María de Maeztu Program for Units of Excellence in R\&D (Grant No. CEX2018-000805-M), Comunidad Autónoma de Madrid through the Nanomag COST-CM Program (Grant No. S2018/NMT-4321), Generalitat Valenciana through Programa Prometeo (2021/017), Centro de Computación Científica of the Universidad Autónoma de Madrid, and Red Española de Supercomputación. 
\end{acknowledgement}

The authors declare no competing financial interest.




\bibliography{ShiftGS} 

\end{document}







\section{\texttt{CRYSTAL} input files \& band structures}
\subsection{MoS$_{2}$}
\begin{figure}[h!]
\centering 
\includegraphics[width=0.7\textwidth]{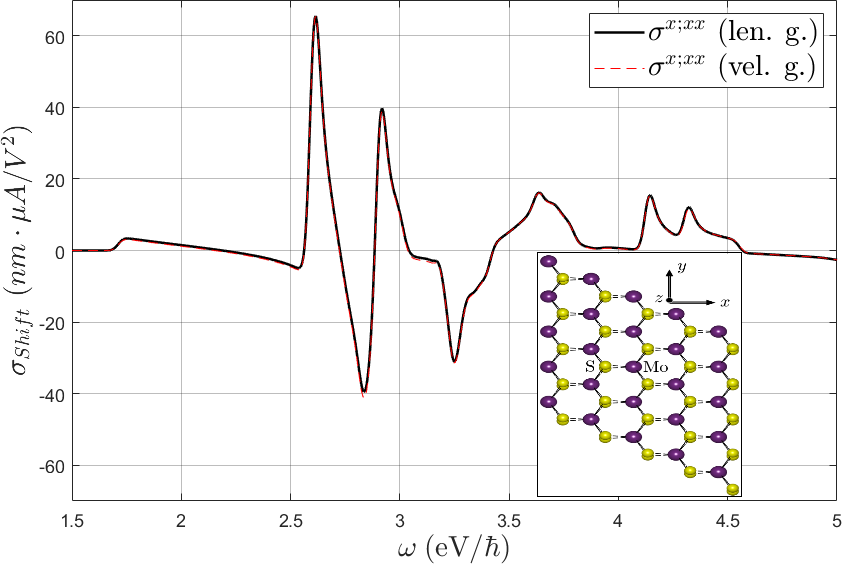}
\end{figure}
\noindent
\texttt{SLAB                   \newline   
69                              \newline                   
3.184                        \newline                      
3                          \newline                         
16 0.66666666667 0.33333333333  1.56355632             \newline   
16 0.66666666667 0.33333333333 -1.56355632             \newline   
242 0.33333333333 0.66666666667 0.             \newline                           
END      \newline
16 19    \newline
0 0 10 2.0 1.0    \newline
1273410.9023000              0.11767088246D-04    \newline
 190697.8300700              0.91478610166D-04    \newline
  43397.8853300              0.48090078640D-03    \newline
  12291.8096770              0.20257193592D-02    \newline
   4009.7420824              0.73190096406D-02    \newline
   1447.3531030              0.23300499900D-01    \newline
    564.30102913             0.65386213610D-01    \newline
    233.74506243             0.15614449910    \newline
    101.56402814             0.29318563787    \newline
     45.805907187            0.36287914289    \newline
0 0 3 2.0 1.0    \newline
    394.27281503             0.18753305081D-01    \newline
    121.72249591             0.16870726663    \newline
     46.754125963            0.63806830653    \newline
0 0 1 2.0 1.0    \newline
     20.923008254            1.0000000    \newline
0 0 1 0.0 1.0    \newline
      8.2685567800           1.0000000    \newline
0 0 1 0.0 1.0    \newline
      3.8629345671           1.0000000    \newline
0 0 1 0.0 1.0    \newline
      1.7794684781           1.0000000    \newline
0 0 1 0.0 1.0    \newline
      0.61064260103          1.0000000    \newline
0 0 1 0.0 1.0    \newline
      0.27412269445          1.0000000    \newline
0 0 1 0.0 1.0    \newline
      0.11325939107          1.0000000    \newline
0 2 8 6.0 1.0    \newline
   2189.8930459              0.23912552864D-03    \newline
    518.94596592             0.20772032158D-02    \newline
    168.19560151             0.11242420571D-01    \newline   
     63.745282788            0.44069933941D-01    \newline
     26.597033077            0.12918778608    \newline
     11.774251449            0.26910820167    \newline
      5.3534379024           0.37855928620    \newline
      2.4701911802           0.29692134655    \newline
0 2 2 4.0 1.0    \newline
     82.120288349           -0.39420318847D-01    \newline
      4.9523532869           0.64048403090    \newline
0 2 1 0.0 1.0    \newline
      1.0828262029           1.0000000    \newline
0 2 1 0.0 1.0    \newline   
      0.49271277356          1.0000000    \newline
0 2 1 0.0 1.0    \newline
      0.20483450942          1.0000000    \newline
0 2 1 0.0 1.0    \newline
      0.80743615716D-01      1.0000000    \newline
0 3 1 0.0 1.0    \newline
      4.15900000             1.0000000    \newline
0 3 1 0.0 1.0    \newline
      1.01900000             1.0000000    \newline
0 3 1 0.0 1.0    \newline    
      0.464000000            1.0000000    \newline
0 3 1 0.0 1.0    \newline
      0.194000000            1.0000000    \newline
242 12    \newline
INPUT    \newline
14. 0 2 4 4 2 0    \newline
  10.097000 180.076853 0    \newline
   4.375670  24.715920 0    \newline
   9.126564  41.227678 0    \newline
   8.863223  82.452670 0    \newline
   4.044948   6.345092 0    \newline
   3.866657  12.458423 0    \newline
   7.535754  19.308744 0    \newline
   7.278976  28.977674 0    \newline
   2.763205   3.189516 0    \newline
   2.772085   4.700169 0    \newline
   6.306633  -7.178888 0    \newline
   6.356448  -9.745978 0    \newline
0 0 2 2 1.0    \newline
  14.0000000000     -0.224900434060    \newline
  12.5000000000      0.331512485550    \newline
0 0 1 2 1.0    \newline
  4.25053415000      1.000000000000    \newline
0 0 1 0 1.0    \newline
  0.65151308000      1.000000000000    \newline
0 0 1 0 1.0    \newline
  0.18122290000      1.000000000000    \newline
0 2 4 6 1.0    \newline
  8.89311179150      0.069994449475    \newline
  5.46891122700     -0.235471418830    \newline
  1.35484730070      0.463154600070    \newline
  0.65494867461      0.488201847100    \newline
0 2 1 0 1.0    \newline
  0.46348506000      1.000000000000    \newline
0 2 1 0 1.0    \newline
  0.24987406000      1.000000000000    \newline
0 2 1 0 1.0    \newline
  0.1                1.000000000000    \newline
0 3 3 4 1.0    \newline
  5.00444454970     -0.021587364862    \newline
  1.77368233240      0.209586800860    \newline
  0.76950591696      0.437308805990    \newline
0 3 1 0 1.0    \newline
  0.56023361000      1.000000000000    \newline    
0 3 1 0 1.0    \newline
  0.20486424000      1.000000000000    \newline
0 4 1 0 1.0    \newline
  0.55911598000      1.000000000000    \newline
99 0    \newline
PRINT    \newline
END    \newline
DFT         \newline
XXLGRID                 \newline                 
PBEXC                               \newline       
ENDDFT                                        \newline
TOLINTEG                                    \newline
10 10 10 10 30    \newline
SHRINK                        \newline              
24 24 24                                  \newline
MAXCYCLE                                    \newline
300    \newline
FMIXING            \newline                       
90    \newline
TOLDEE            \newline                         
12                 \newline
NODIIS                \newline
END         }

\newpage
\subsection{GeS}
\begin{figure}[h!]
\centering 
\includegraphics[width=0.7\textwidth]{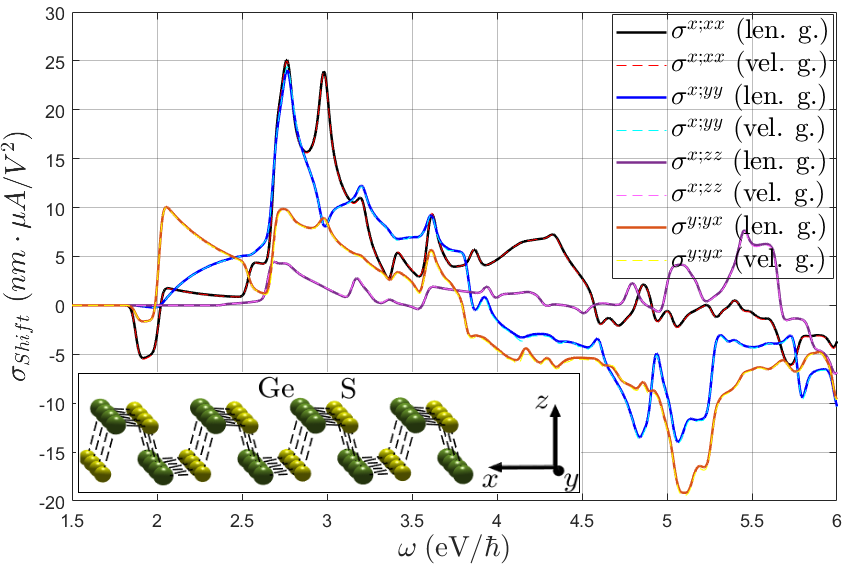}
\end{figure}
\noindent
\texttt{SLAB    \newline
31                          \newline            
4.47   3.66                    \newline                
2          \newline
16   0.000000   0.000000   1.07        \newline               
32  -0.131991   0.000000  -1.28   \newline                                     
END     \newline
16 19   \newline
0 0 10 2.0 1.0   \newline  
1273410.9023000              0.11767088246D-04   \newline
 190697.8300700              0.91478610166D-04   \newline
  43397.8853300              0.48090078640D-03   \newline
  12291.8096770              0.20257193592D-02   \newline
   4009.7420824              0.73190096406D-02   \newline
   1447.3531030              0.23300499900D-01   \newline
    564.30102913             0.65386213610D-01   \newline
    233.74506243             0.15614449910   \newline
    101.56402814             0.29318563787   \newline
     45.805907187            0.36287914289   \newline
0 0 3 2.0 1.0   \newline
    394.27281503             0.18753305081D-01   \newline
    121.72249591             0.16870726663   \newline
     46.754125963            0.63806830653   \newline
0 0 1 2.0 1.0   \newline
     20.923008254            1.0000000   \newline
0 0 1 0.0 1.0   \newline
      8.2685567800           1.0000000   \newline
0 0 1 0.0 1.0   \newline
      3.8629345671           1.0000000   \newline
0 0 1 0.0 1.0   \newline
      1.7794684781           1.0000000   \newline
0 0 1 0.0 1.0   \newline
      0.61064260103          1.0000000   \newline
0 0 1 0.0 1.0   \newline
      0.27412269445          1.0000000   \newline
0 0 1 0.0 1.0   \newline
      0.11325939107          1.0000000   \newline
0 2 8 6.0 1.0   \newline
   2189.8930459              0.23912552864D-03   \newline
    518.94596592             0.20772032158D-02   \newline
    168.19560151             0.11242420571D-01   \newline
     63.745282788            0.44069933941D-01   \newline
     26.597033077            0.12918778608   \newline
     11.774251449            0.26910820167   \newline
      5.3534379024           0.37855928620   \newline
      2.4701911802           0.29692134655   \newline
0 2 2 4.0 1.0   \newline
     82.120288349           -0.39420318847D-01   \newline
      4.9523532869           0.64048403090   \newline
0 2 1 0.0 1.0   \newline
      1.0828262029           1.0000000   \newline
0 2 1 0.0 1.0   \newline   
      0.49271277356          1.0000000   \newline
0 2 1 0.0 1.0   \newline
      0.20483450942          1.0000000   \newline
0 2 1 0.0 1.0   \newline
      0.80743615716D-01      1.0000000   \newline
0 3 1 0.0 1.0   \newline
      4.15900000             1.0000000   \newline
0 3 1 0.0 1.0   \newline
      1.01900000             1.0000000   \newline
0 3 1 0.0 1.0   \newline
      0.464000000            1.0000000   \newline
0 3 1 0.0 1.0   \newline
      0.194000000            1.0000000   \newline
32 22   \newline
0 0 11 2.0 1.0   \newline
7233056.0346000              0.76638751457D-05   \newline
1082886.1731000              0.59603601369D-04   \newline
 246481.4695900              0.31319088031D-03   \newline
  69862.4269550              0.13194051561D-02   \newline
  22815.8096620              0.47736099191D-02   \newline
   8246.5369297              0.15319250467D-01   \newline
   3219.9367257              0.43900870673D-01   \newline
   1336.5743706              0.11028573632   \newline
    582.87737501             0.22912630489   \newline
    264.59511360             0.34779259246   \newline
    123.77823320             0.29968722223   \newline
0 0 4 2.0 1.0   \newline
   2311.1055804              0.75033084731D-02   \newline
    716.27089868             0.74778386626D-01   \newline
    275.45330910             0.35092882302   \newline
    118.93292565             0.72055989686   \newline
0 0 1 2.0 1.0   \newline
     58.435699085            1.0000000   \newline
0 0 1 2.0 1.0   \newline
     26.261575973            1.0000000   \newline
0 0 1 0.0 1.0   \newline
     12.664880671            1.0000000    \newline
0 0 1 0.0 1.0   \newline
      5.7269548505           1.0000000   \newline
0 0 1 0.0 1.0   \newline
      2.7555023203           1.0000000   \newline
0 0 1 0.0 1.0   \newline
      1.2432886754           1.0000000   \newline
0 0 1 0.0 1.0   \newline
      0.33107183846          1.0000000   \newline
0 0 1 0.0 1.0   \newline
      0.15964081368          1.0000000   \newline
0 0 1 0.0 1.0   \newline
      0.68463178923D-01      1.0000000   \newline
0 2 9 6.0 1.0   \newline
  16555.7110740              0.10199370391D-03   \newline
   3914.9903745              0.90504190546D-03   \newline
   1269.4766518              0.51558490762D-02   \newline
    484.35437789             0.22010335708D-01   \newline
    204.64985822             0.73434798827D-01   \newline
     92.791032094            0.18608955294   \newline
     44.055060255            0.33330981945   \newline
     21.601319931            0.36229776246   \newline
     10.732169233            0.17765186263   \newline
0 2 2 6.0 1.0   \newline
    100.88933962            -0.11146182144   \newline
     36.640254741           -0.52324324314   \newline
0 2 4 2.0 1.0   \newline
      8.0343224115           0.31484763263   \newline
      4.5613438333           0.38737744790   \newline
      3.1794882223           0.50034353638   \newline
      0.92612041457          0.22656707948   \newline
0 2 1 0.0 1.0   \newline
      1.7977040907           1.0000000   \newline
0 2 1 0.0 1.0   \newline
      0.41386865432          1.0000000   \newline
0 2 1 0.0 1.0   \newline
      0.15830798709          1.0000000   \newline
0 2 1 0.0 1.0   \newline
      0.57877035911D-01      1.0000000   \newline
0 3 7 10.0 1.0   \newline
    420.99565921             0.78212192196D-03   \newline
    126.36209818             0.69560123831D-02   \newline
     48.661473548            0.33712188323D-01   \newline
     20.880325527            0.10717291194   \newline
      9.5418229355           0.23161030141   \newline
      4.4353653868           0.33390954301   \newline
      2.0285421942           0.33720676378   \newline
0 3 1 0.0 1.0   \newline
      0.89484935267          0.21702473265    \newline
0 3 1 0.0 1.0   \newline
      0.35786074334          1.0000000   \newline
0 3 1 0.0 1.0   \newline
      0.1400000              1.0000000   \newline
99  0       \newline
PRINT   \newline
END   \newline
DFT         \newline
XXLGRID                \newline                 
PBEXC                             \newline      
END                                        \newline
TOLINTEG                                   \newline
10 10 10 10 30   \newline
SHRINK                      \newline               
24 24 24                               \newline   
MAXCYCLE                                   \newline
300   \newline
FMIXING          \newline                        
40   \newline
TOLDEE          \newline                          
12                         \newline               
END   }

\newpage
\subsection{GaAs}
\begin{figure}[h!]
\centering 
\includegraphics[width=0.7\textwidth]{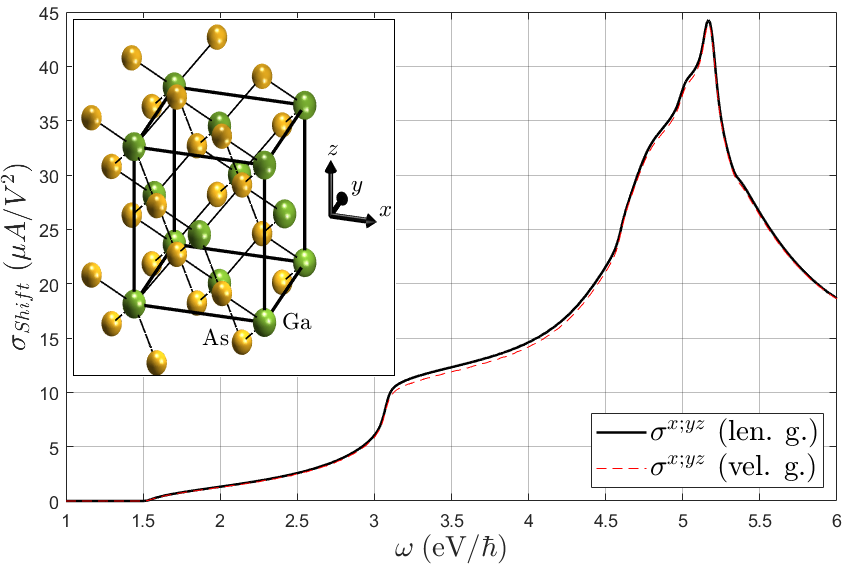}
\end{figure}
\noindent
\texttt{CRYSTAL    \newline
0 0 0              \newline                      
216                   \newline                     
5.65                   \newline                
2                             \newline             
231    0.     0.    0.    \newline
233    0.25   0.75  0.75       \newline                    
END      \newline
231 9    \newline
INPUT    \newline
21. 0 2 4 6 2 0    \newline
 25.880361 370.273040 0    \newline
 7.901295 9.190615 0    \newline
 45.149190 99.144001 0    \newline
 44.979981 198.295512 0    \newline
 17.224251 28.445653 0    \newline
 16.747329 56.949705 0    \newline
 51.968812 -18.168797 0    \newline
 51.629117 -27.380273 0    \newline
 15.241738 -1.587022 0    \newline
 15.320193 -2.516292 0    \newline
 4.918589 0.083166 0    \newline
 4.755103 0.202198 0    \newline
 10.762263 -0.616990 0    \newline
 19.852939 -3.138584 0    \newline
0 0 6 2. 1.    \newline
       2848.20000         0.362000000E-03    \newline
       420.664000         0.211700000E-02    \newline
       29.8118000         0.118964000    \newline
       14.2207000        -0.461723000    \newline
       2.67643000         0.751559000    \newline
       1.13353000         0.447202000    \newline
0 0 6 2. 1.    \newline
       2848.20000        -0.970000000E-04    \newline
       420.664000        -0.614000000E-03    \newline
       29.8118000        -0.310690000E-01    \newline
       14.2207000         0.126784000    \newline
       2.67643000        -0.264288000    \newline
       1.13353000        -0.275471000    \newline
0 0 1 0. 1.    \newline
      0.207220000          1.00000000    \newline
0 0 1 0. 1.    \newline
      0.120000000         1.00000000    \newline
0 2 6 6. 1.    \newline
       109.624000         0.210100000E-02    \newline
       21.0855000        -0.801960000E-01    \newline
       4.92260000         0.396415000    \newline
       2.15591000         0.519076000    \newline
      0.901913000         0.207520000    \newline
      0.202004000         0.782500000E-02    \newline
0 2 6 1. 1.    \newline
       109.624000        -0.288000000E-03    \newline
       21.0855000         0.135550000E-01    \newline
       4.92260000        -0.736290000E-01    \newline
       2.15591000        -0.120860000    \newline
      0.901913000        -0.196000000E-02    \newline
      0.202004000         0.493206000    \newline
0 2 1 0. 1.    \newline
      0.120000000         1.00000000    \newline
0 3 6 10. 1.    \newline
       85.7978000         0.146680000E-01    \newline
       27.6822000         0.856210000E-01    \newline
       10.1760000         0.248336000    \newline
       3.92208000         0.401414000    \newline
       1.45858000         0.398604000    \newline
      0.488760000         0.186898000    \newline
0 3 1 0. 1.    \newline
      0.177200000          1.00000000    \newline
233 9    \newline
INPUT    \newline
23. 0 2 4 6 2 0    \newline
 28.725122 370.114025 0    \newline
 6.767681 9.349296 0    \newline
 45.331064 99.142103 0    \newline
 44.767415 198.307880 0    \newline
 19.539090 28.383073 0    \newline
 18.973471 56.871464 0    \newline
 51.057152 -18.485145 0    \newline
 50.151340 -28.113530 0    \newline
 16.108936 -1.223895 0    \newline
 14.672223 -1.345765 0    \newline
 3.851927 0.101757 0    \newline
 3.813502 0.170338 0    \newline
 11.940584 -0.775230 0    \newline
 17.761160 -2.157259 0    \newline
0 0 6 2. 1.    \newline
       2542.81000         0.113700000E-02    \newline
       381.169000         0.605500000E-02    \newline
       40.2342000         0.841250000E-01    \newline
       16.1217000        -0.405285000    \newline
       3.20189000         0.712926000    \newline
       1.42096000         0.473376000    \newline
0 0 6 2. 1.    \newline
       2542.81000        -0.390000000E-03    \newline
       381.169000        -0.219000000E-02    \newline
       40.2342000        -0.268530000E-01    \newline
       16.1217000         0.136878000    \newline
       3.20189000        -0.320457000    \newline
       1.42096000        -0.337391000    \newline
0 0 1 0. 1.    \newline
      0.321443000          1.00000000    \newline
0 0 1 0. 1.    \newline
      0.120000000          1.00000000    \newline
0 2 6 6. 1.    \newline
       99.5349000         0.385700000E-02    \newline
       24.1195000        -0.851010000E-01    \newline
       5.84196000         0.404762000    \newline
       2.56010000         0.531478000    \newline
       1.09308000         0.184012000    \newline
      0.318424000         0.576400000E-02    \newline
0 2 6 3. 1.    \newline
       99.5349000        -0.772000000E-03    \newline
       24.1195000         0.199410000E-01    \newline
       5.84196000        -0.107210000    \newline
       2.56010000        -0.172259000    \newline
       1.09308000         0.876100000E-02    \newline
      0.318424000         0.569744000    \newline
0 2 1 0. 1.    \newline
      0.120000000          1.00000000    \newline    
0 3 6 10. 1.    \newline
       113.509000         0.119800000E-01    \newline
       36.8872000         0.795440000E-01    \newline
       13.6893000         0.236755000    \newline
       5.38964000         0.401534000    \newline    
       2.08046000         0.406686000    \newline
      0.737568000         0.173162000    \newline
0 3 1 0. 1.    \newline
      0.307800000          1.00000000    \newline
99  0        \newline
PRINT    \newline
END    \newline
DFT        \newline
XXLGRID                \newline                  
HSE06                              \newline   
END                                         \newline
TOLINTEG                                    \newline
10 10 10 10 30    \newline
SHRINK                        \newline              
24 24 24                                  \newline 
MAXCYCLE                                    \newline
300    \newline
FMIXING            \newline                       
40    \newline
TOLDEE            \newline                         
12                            \newline         
END  } 

\newpage
\subsection{BaTiO$_{3}$}
\begin{figure}[h!]
\centering 
\includegraphics[width=0.7\textwidth]{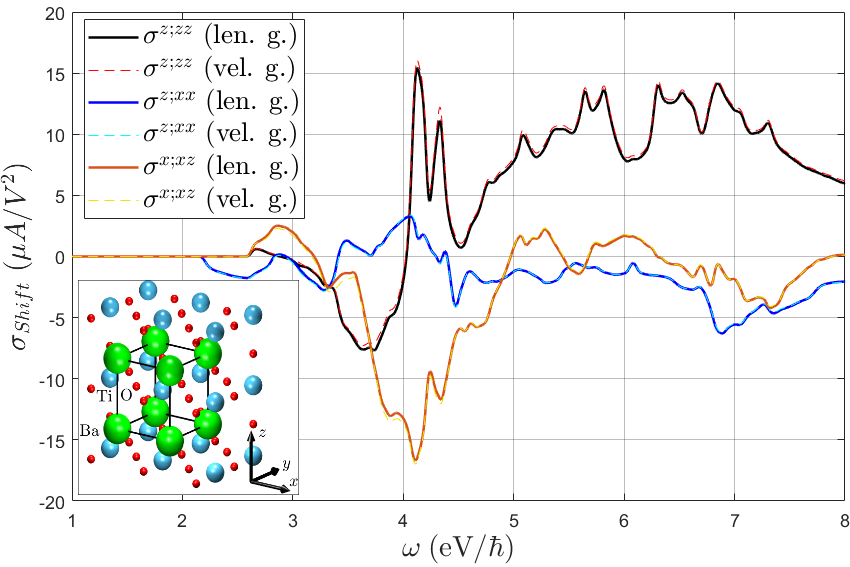}
\end{figure}
\noindent
\texttt{CRYSTAL   \newline
0 0 0              \newline                  
99                     \newline                
3.99  4.1                 \newline              
4                            \newline          
8    0.   0.   0.548972     \newline
8    0.   0.5  0.064221   \newline
22   0.   0.   0.100043   \newline 
256  0.5  0.5  0.582541    \newline
END   \newline
8 12    \newline
0 0 8 2.0 1.0
 116506.4690800              0.40383857939D-04    \newline
  17504.3497240              0.31255139004D-03    \newline
   3993.4513230              0.16341473495D-02    \newline
   1133.0063186              0.68283224757D-02  \newline
    369.99569594             0.24124410221D-01    \newline
    133.62074349             0.72730206154D-01    \newline
     52.035643649            0.17934429892    \newline
     21.461939313            0.33059588895    \newline
0 0 2 2.0 1.0    \newline
     89.835051252            0.96468652996D-01    \newline
     26.428010844            0.94117481120    \newline
0 0 1 0.0 1.0    \newline
      9.2822824649           1.0000000    \newline
0 0 1 0.0 1.0    \newline
      4.0947728533           1.0000000    \newline
0 0 1 0.0 1.0    \newline
      1.3255349078           1.0000000    \newline
0 0 1 0.0 1.0    \newline
      0.51877230787          1.0000000    \newline
0 0 1 0.0 1.0    \newline
      0.19772676454          1.0000000    \newline
0 2 5 4.0 1.0    \newline
    191.15255810             0.25115697705D-02    \newline
     45.233356739            0.20039240864D-01    \newline
     14.353465922            0.93609064762D-01    \newline
      5.2422371832           0.30618127124    \newline
      2.0792418599           0.67810501439    \newline
0 2 1 0.0 1.0    \newline
      0.84282371424          1.0000000    \newline
0 2 1 0.0 1.0    \newline
      0.33617694891          1.0000000    \newline
0 2 1 0.0 1.0    \newline
      0.12863997974          1.0000000    \newline
0 3 1 0.0 1.0    \newline
  0.4534621300      1.0000000000000    \newline
22 18    \newline
0 0 11 2.0 1.0    \newline
3070548.8651000              0.86954016630D-05    \newline
 460777.8864300              0.67452737727D-04    \newline
 104901.2288900              0.35477293028D-03    \newline
  29695.8611990              0.14977525588D-02    \newline
   9678.8892688              0.54309912055D-02    \newline
   3490.1877912              0.17439360524D-01    \newline
   1359.2217621              0.49835634640D-01    \newline
    562.42721208             0.12379633943    \newline
    244.22296250             0.25057490943    \newline
    110.16668710             0.35934609007    \newline
     50.881903357            0.27594242664    \newline
0 0 4 2.0 1.0    \newline
    965.95430789             0.41773927781D-02    \newline
    299.27072059             0.40277148567D-01    \newline
    114.83772939             0.17898686817    \newline
     49.477578954            0.31783043543    \newline
0 0 1 2.0 1.0    \newline
     22.982839977            1.0000000    \newline
0 0 1 2.0 1.0    \newline
     10.518305037            1.0000000    \newline
0 0 1 0.0 1.0    \newline
      4.9774390567           1.0000000    \newline
0 0 1 0.0 1.0    \newline
      2.1339846838           1.0000000    \newline
0 0 1 0.0 1.0    \newline
      1.0342457284           1.0000000    \newline
0 0 1 0.0 1.0    \newline
      0.46199774995          1.0000000    \newline
0 0 1 0.0 1.0    \newline
      0.10621264194          1.0000000    \newline
0 2 9 6.0 1.0    \newline
   5169.6755427              0.18802523596D-03    \newline
   1225.0961638              0.16473458826D-02    \newline
    397.60051934             0.91104554321D-02    \newline
    151.36154684             0.36987450926D-01    \newline
     63.613321773            0.11376329624    \newline
     28.514560307            0.25345208574    \newline
     13.248003298            0.38019402704    \newline
      6.3048807760           0.30989136346    \newline
      2.9493525821           0.87418944007D-01    \newline
0 2 5 6.0 1.0    \newline
     40.738772213           -0.73233793267D-02    \newline
     14.062358461           -0.34282591082D-01    \newline
      2.7460680961           0.35655009250    \newline
      1.2713688141           0.76112035427    \newline
      0.57610015545          0.62716743237    \newline
0 2 1 0.0 1    \newline
      0.23981820724          1.0000000    \newline
0 2 1 0.0 1.0    \newline
      0.11000000000          1.0000000    \newline
0 3 5 2.0 1.0    \newline
     89.589880075            0.21223030030D-02    \newline
     26.591412960            0.15911819913D-01    \newline
      9.7739715702           0.62875243121D-01    \newline
      3.9625083655           0.17144170807    \newline
      1.6890532654           0.30565506624    \newline
0 3 1 0.0 1.0    \newline
      0.71539771469          1.0000000    \newline
0 3 1 0.0 1.0    \newline
      0.29366677748          1.0000000    \newline
0 3 1 0.0 1.0    \newline
      0.11079094851          1.0000000    \newline
0 4 1 0.0 1.0    \newline
  0.56200000000      1.00000000000000    \newline
256 10    \newline
INPUT    \newline
10. 0 2 4 4 4 2    \newline
  4.177931587  84.785457583 0    \newline
  2.522632800  17.372709041 0    \newline
  6.294119351  52.512225743 0    \newline
  6.476457746 105.022668647 0    \newline
  2.284326647   8.707014937 0    \newline
  2.091555201  17.165458832 0    \newline
  1.925291745   5.346535679 0    \newline
  1.878534118   8.025720742 0    \newline
  0.907088727   1.346295081 0    \newline
  0.910060953   2.063710453 0    \newline
  6.256321669 -20.003223472 0    \newline
  6.134135837 -26.118214748 0    \newline
  1.641382784  -2.344457989 0    \newline
  1.599343316  -2.980867480 0    \newline
  2.142381001  -3.316602759 0    \newline
  2.159981109  -4.275647018 0    \newline
0 0 2 2 1.0    \newline
      5.7000000000           1.3292982002    \newline
      5.2043612469          -1.7367497134    \newline
0 0 1 2 1.0    \newline
      1.955486070000       1.000000000000    \newline
0 0 1 0 1.0    \newline
      0.41033660537          1.0000000    \newline
0 0 1 0 1.0    \newline
      0.18393532770          1.0000000    \newline
0 0 1 0 1.0    \newline
  0.72521216545D-01      1.0000000    \newline
0 2 4 6 1.0    \newline
      5.6000000000           0.51385026013    \newline
      5.1104879274          -0.70529238864    \newline
      2.5006544803           0.42822329830    \newline
      0.51947995345         -0.57125984472    \newline
0 2 1 0 1.0    \newline
      0.24536994841          1.0000000    \newline
0 2 1 0 1.0    \newline
      0.11256056700          1.0000000    \newline
0 3 3 0 1.0    \newline
      2.7000000000            .14372069823827    \newline   
      2.3460740881           -.19565347309413    \newline
      0.39553542609           .29963058551591    \newline
0 3 1 0 1.0    \newline
      0.12967935859          1.0000000    \newline
99  0        \newline
PRINT    \newline
END    \newline
DFT       \newline
XXLGRID               \newline                    
PBEXC                             \newline   
END                                         \newline
TOLINTEG                                    \newline
8 8 8 10 20    \newline
SHRINK                     \newline                 
24 24 24                               \newline    
MAXCYCLE                                    \newline
500    \newline
FMIXING            \newline                       
95    \newline
TOLDEE            \newline                         
12        \newline
GUESSP                \newline                 
END      }
\newline 
\\
\vspace{0.1cm} \noindent
Note: the calculation was restarted from a previous iteration with MAXCYCLE$=500$ and FMIXING$=90$.

\newpage
\section{Comparison of BZ \& IBZ summations}
\begin{figure}[h!]
\centering 
\includegraphics[width=0.7\textwidth]{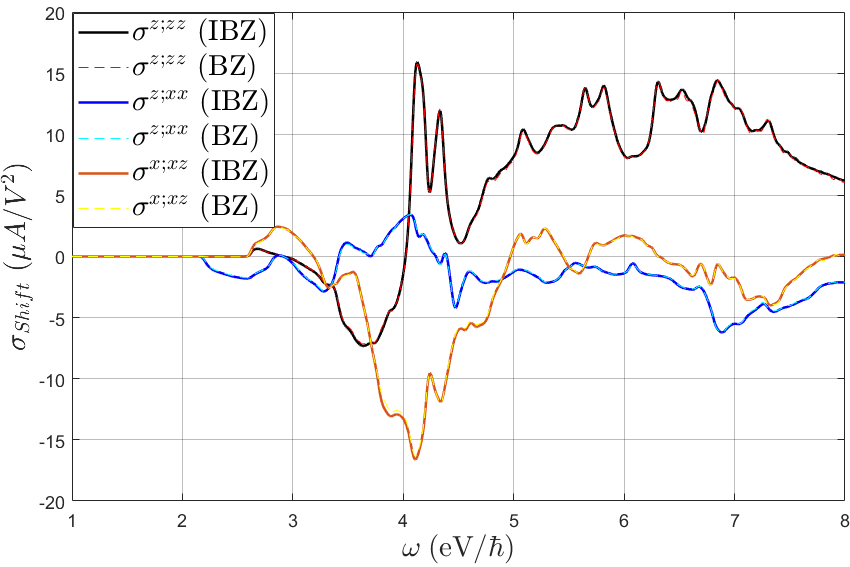}
\caption{Shift conductivity tensor in BaTiO$_{3}$ computed in the velocity gauge with the standard BZ (dashed lines) and folded IBZ summations with time-reversal symmetry (solid lines). The BZ grid dimension was $100\times100\times100$, from which the IBZ grid was obtained by restriction as specified in the main text (other computational parameters as therein). The computation time ratio $t_{BZ}/t_{IBZ}$ between the Brillouin zone (BZ) and irreducible Brillouin zone (IBZ, or representation domain) summations in identical computational conditions was $15.94$, which is approximately $\text{Volume BZ}/\text{Volume IBZ}=2\abs{F}=\abs{C_{4v}+\pazocal{T}C_{4v}}=16$. }
\end{figure}

